\documentclass[a4paper]{article}
\usepackage{amsmath,amssymb}
\usepackage{ifpdf}
\ifpdf
  \usepackage[pdftex]{graphicx}
\else
 \usepackage[dvipdfm]{graphicx}
\fi
\makeatletter
    
    \@addtoreset{equation}{section}
  \makeatother

\newcommand{\eref}[1]{(\ref{#1})}

\newsavebox{\boxa}

\begin{document}

%%%%%%%%%%%%%%%%%%%%%%%%%%%%%%%%%%%%%%%%%%%%
\thispagestyle{empty}
\begin{flushright}
OU-HET 736
\end{flushright}
\vskip3cm
\begin{center}
{\Large {\bf Restoration of Lorentz Symmetry\\for Lifshitz-Type Scalar Theory }}
\vskip1.5cm
{\large 
{Kengo Kikuchi\footnote{kikuchi@het.phys.sci.osaka-u.ac.jp}
}

}
\vskip.5cm
{\it Department of Physics, Graduate School of Science, 
\\
Osaka University, Toyonaka, 560-0043, Japan}
\end{center}

%%%%%%%%%%%%%%%%%%%%%%%%%%%%%%%%%%%%%%%%%%%%
\vskip2cm
\begin{abstract}
The purpose of this paper is to present our study on the restoration of the Lorentz symmetry for a Lifshitz-type scalar theory in the infrared region by using nonperturbative methods. We apply the Wegner-Houghton equation, which is one of the exact renormalization group equations, to the Lifshitz-type theory. Analyzing the equation for a $z=2, d=3+1$ Lifshitz-type scalar model, and using some variable transformations, we found that broken symmetry terms vanish in the infrared region. 
This shows that the Lifshitz-type scalar model dynamically restores the Lorentz symmetry at low energy. Our result provides a definition of ultraviolet complete renormalizable scalar field theories. These theories can have nontrivial interaction terms of $\phi^{n} (n=4, 6, 8, 10)$ even when the Lorentz symmetry is restored at low energy.
\end{abstract}

%%%%%%%%%%%%%%%%%%%%%%%%%%%%%%%%%%%%%%%%%%

\newpage
\section{ Introduction}
 Recently, Lifshitz-type theory~\cite{EMLifshitz} has been the focus of much attention~\cite{Horava:2009uw}. This theory has an anisotropic scaling for space and time at the Lifshitz fixed point.  In this theory, we substitute the second-order space differential operator in the kinetic term in the action with the 2$z$ order one as follows:
 \begin{eqnarray}
S &=& \int dtd^Dx    
\frac{1}{2} \{ \phi (-\partial_0 \partial_0 + (-\partial_i \partial_i )^z+m^{2z} )\phi \}\nonumber\\
&=& \int_{p, p'}  \frac{1}{2}  (p_0^2+ p_i^{2z}+m^{2z} )\phi_p\phi_{p'}\delta(p+p').
\end{eqnarray}
It is found from this equation that the time dimension is $z$, while the space dimension is one. The advantage of the Lifshitz-type theory is that the higher derivative terms in the kinetic terms suppress the ultraviolet (UV) divergence. This feature broadens the class of perturbatively renormalizable field theories.

As a compensation for these good UV properties, one has to sacrifice the Lorentz symmetry in the UV region. If the Lifshitz-type theory indeed explains physical phenomena at our energy scale properly, the theory should restore the symmetry in the infrared (IR) region. References~\cite{Iengo:2009ix, Anselmi:2009vz, Gomes:2011pq, Anselmi:2008bq, Anselmi:2008bs} are related to the Lorentz symmetry in the Lifshitz-type theory in the IR region. It is well known that, from naive power counting, it is expected that the symmetry can be restored in the IR region. However, the restoration of the symmetry should also be examined nonperturbatively. The goal of this paper is to study whether the theory defined at the Lifshitz fixed point really flows into the Lorentz invariant theory at low energy by using the exact renormalization group equation. Such a problem is also discussed in a large $N$ limit in Refs.~\cite{Dhar:2009dx, Dhar:2009am}.

The exact renormalization group equation as typified in Ref.~\cite{Wilson:1973jj} enables us to analyze theories nonperturbatively. For example, Ref.~\cite{Bervillier:2004mh} discussed the Lifshitz-type theory using the Wilson-Polchinski exact renormalization group equation. In our work, we apply the Wegner-Houghton equation, which is one of the exact renormalization group equations, to the Lifshitz-type theory, and analyze the behavior of the theory space of the Lifshitz-type theory. It is found that this theory has a Lorentz symmetrical Gaussian IR fixed point, and we confirm that the method in this paper indeed reproduces the previously mentioned naive power counting arguments at the leading order in the perturbation theory. Using numerical analysis, we find that symmetry-violating terms in the theory vanish in the IR region. In conclusion, the $z=2, d=3+1$ Lifshitz-type scalar model restores the Lorentz symmetry in the IR region. Our result provides a definition of ultraviolet complete renormalizable scalar field theories. Remarkably, these theories can have nontrivial interaction terms of $\phi^{n} (n=4, 6, 8, 10)$ even when the Lorentz symmetry is restored at low energy.

This paper is organized as follows. In section \ref{WH}, we use a cutoff method to apply the Wegner-Houghton equation to the Lifshitz-type theory. In section \ref{MA}, we introduce our model and apply the equation derived in section \ref{WH} to this model, and in section \ref{NA}, we analyze the equations numerically.  Finally, we give a summary in section \ref{SD}. Throughout the paper, we work on the Euclidean theory, and we give our notation in Appendix \ref{appa}.

\section{Wegner-Houghton Equation for Lifshitz-Type Theory}\label{WH}
The Wegner-Houghton equation is an exact renormalization group equation~\cite{Wegner:1972ih}. We review the derivation of the equation in Appendix \ref{appb} following Refs.~\cite{Clark:1992jr,Aoki:2000wm}. The equation for the effective action $S$ is
\begin{eqnarray}
\Lambda \frac{d}{d\Lambda} S &=& -\frac{1}{2\delta t} \Bigl\{ \mathrm{tr} \ln\Bigl(\frac{\delta^2 S}{\delta \Omega \delta \Omega}\Bigl)-\frac{\delta S}{\delta \Omega} \Bigl(\frac{\delta^2 S}{\delta \Omega \delta \Omega}\Bigl)^{-1}  \frac{\delta S}{\delta \Omega} \Bigl\}\nonumber\\
&&-dS+\int_{p}\Omega_{p}^{i}\Bigl(d_{\Omega}-\gamma+{\hat{p}^{\mu}} \frac{\partial'}{\partial {\hat{p}^{\mu}}}\Bigl) \frac{\delta}{\delta \Omega_{p}^{i}}S,\label{eq00}
\end{eqnarray}
where $\Lambda$ is a cutoff, $\Omega$ is a general field, i.e., $\Omega=\phi$ in the case of a real scalar field, $\gamma$ is an anomalous dimension, and $d_\Omega$ is the dimension of the field.  The definitions of $\delta t, \hat{p},$ and $\partial'$ are given in Appendix \ref{appb}. The first term on the RHS is the contribution from shell-mode integrals, and the second and third terms are from the scaling part. When we discuss the Lifshitz-type theory, $d$ in Eq.~\eref{eq00} is rewritten as a sum of space dimension $D$ and time $z$, that is, $d\rightarrow D+z$.

\begin{figure}[h]\begin{center}\includegraphics [width=70mm]{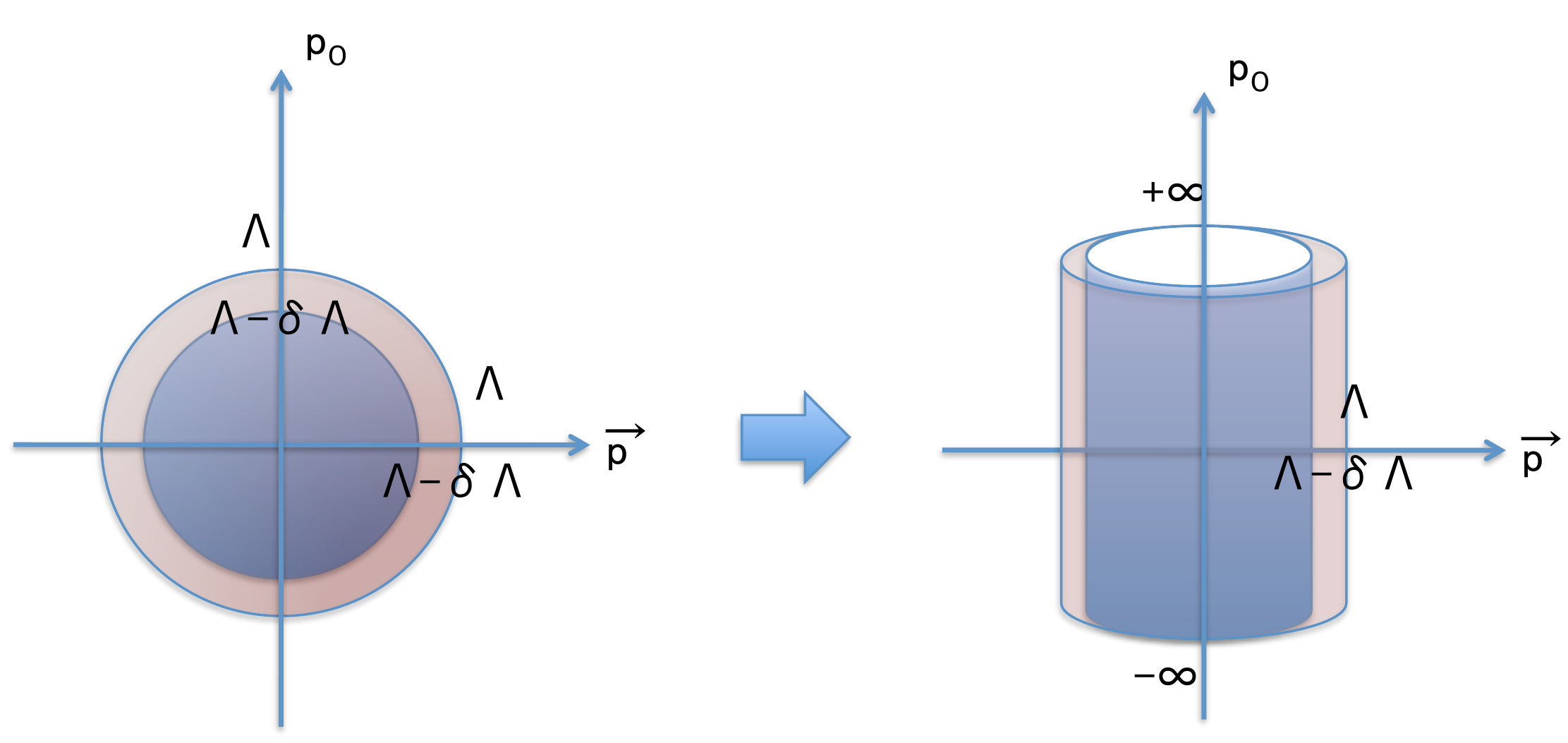}\end{center}
\caption{Cutoff method.  The left side of the figure is the usual cutoff.  The momentum region is a ball inside a sphere in space and time with the radius $\sqrt{p_0^2+p_i^2}=\Lambda$. It has a symmetry between space and time.  The right side is a cylindrical cutoff; $p_0$ is integrated out from $-\infty$ to  $\infty$. }
\label{fig1}
\end{figure}

To solve the equation in the Lifshitz-type theory, we need to know how to perform momentum integrals.  Because the Lifshitz-type theory does not have Lorentz symmetry, it is difficult to understand how to integrate out the shell-mode momentum.  There are various discussions on cutoff methods~\cite{Dhar:2009am,LopezNacir:2011mt}. In this work, we use a cylindrical cutoff as an alternative to a spherical one. (See Fig.~\ref{fig1}.)

\section{Models and Analysis}\label{MA}
\subsection{$z=2, d=3+1$ Lifshitz-Type Scalar Model}
In general, we need to truncate the action to solve the renormalization group equations concretely. Lifshitz-type scalar theories are classified clearly in Refs.~\cite{Iengo:2009ix, Anselmi:2007ri}. In this paper, we adopt an effective action that contains all interactions for which the dimensions of the coupling in units of mass are more than or equal to 0, that is, relevant or marginal operators by naive power counting as a truncation.  We also impose a $\mathbb{Z}_2$ symmetry. The action is
\begin{eqnarray}
S &=& \int _ {x}  \Bigl\{   
\frac{1}{2} (\partial_0 \phi \partial_0 \phi+\beta_0\partial_i \partial_i \phi \partial_j \partial_j \phi+m^4\phi^2 )\nonumber\\
&&+\frac{\lambda_4}{4!} \phi^4 +\frac{\lambda_6}{6!} \phi^6 +\frac{\lambda_8}{8!} \phi^8+\frac{\lambda_{10} }{10!}  \phi^{10} \nonumber\\
&&+ \frac{1}{2} \Bigl(\alpha_0\partial_i \phi \partial_i \phi+\frac{\alpha_2}{2!} \phi^2 \partial_i \phi \partial_i \phi+ \frac{\alpha_4}{4!} \phi^4 \partial_i \phi \partial_i \phi\Bigl) \Bigl\},\label{eq0}
\end{eqnarray}
where the dimensions for $x, t, \phi$, and the parameters in the action are 
\begin{eqnarray}
&&[x]=-1, [t]=-2, [\alpha_0]=2, [\alpha_2]=1, [\alpha_4]=0, [\beta_0]=0,[\phi] = \frac{1}{2},\nonumber\\
&&[m^4]=4, [\lambda_4]=3, [\lambda_6]=2, [\lambda_8]=1, [\lambda_{10}]=0. 
\end{eqnarray}

In terms of the derivative expansion~\cite{Morris:1993qb, Morris:1994ie, Morris:1997xj}, the action in Eq.~\eref{eq0} is the sum of three parts. The first line is the kinetic terms of the free scalar Lifshitz-type theory, and the second and third lines are the local potential approximation terms and first order of the derivative expansion terms, that is,
\begin{eqnarray}
S &=&S_{\mathrm{Lifshitz (free)}}+S_{\mathrm{LPA (int)}}+S_{\mathrm{Diff (int)}}.
\end{eqnarray}
Note that the term $\partial_i \phi \partial_i \phi$, which is needed for Lorentz symmetry, naturally appears in $S_{\mathrm{Diff (int)}}$. To restore the symmetry in the IR region, interaction terms that break the symmetry should vanish in the IR region.

We obtain the Wegner-Houghton equation for the Lifshitz-type theory as one of the main results of this paper. The details of the calculations are given in Appendix \ref{appc}. The Wegner-Houghton equation in the present model reads
{\allowdisplaybreaks \begin{eqnarray}
\frac{\delta\alpha_0}{\delta t}&=&2\alpha_0 +\frac{\alpha_2}{8\pi^2(\alpha_0+\beta_0+m^4)^{1/2}},\label{eq1}\\
\frac{\delta\beta_0}{\delta t}&=&0,\\
\frac{\delta m^4}{\delta t}&=&4m^4 +\frac{1}{8\pi^2(\alpha_0+\beta_0+m^4)^{1/2}}(\alpha_2+\lambda_4),\label{eq1000}\\
\frac{\delta \lambda_4}{\delta t}&=&3\lambda_4 - \frac{1}{16\pi^2(\alpha_0+\beta_0+m^4)^{3/2}}\{3(\alpha_2+\lambda_4)^2\}\nonumber\\
&&+\frac{1}{8\pi^2(\alpha_0+\beta_0+m^4)^{1/2}}(\alpha_4+\lambda_6),\label{eq1001}\\
\frac{\delta \lambda_6}{\delta t}&=&2\lambda_6 + \frac{1}{32\pi^2(\alpha_0+\beta_0+m^4)^{5/2}}\{45(\alpha_2+\lambda_4)^3\}\nonumber\\
&&-\frac{1}{16\pi^2(\alpha_0+\beta_0+m^4)^{3/2}}\{15(\alpha_2+\lambda_4)(\alpha_4+\lambda_6)\} \nonumber \\
&&+\frac{\lambda_8}{8\pi^2 (\alpha_0+\beta_0+m^4)^{1/2}},\\
\frac{\delta \lambda_{8}}{\delta t}&=&\lambda_8 -\frac{1}{64\pi^2(\alpha_0+\beta_0+m^4)^{7/2}}\{1575(\alpha_2+\lambda_4)^4\}\nonumber\\
&&+\frac{1}{16\pi^2(\alpha_0+\beta_0+m^4)^{5/2}}\{315(\alpha_2+\lambda_4)^2(\alpha_4+\lambda_6)\}\nonumber\\
&&-\frac{1}{16\pi^2 (\alpha_0+\beta_0+m^4)^{3/2}}\{35(\alpha_4+\lambda_6)^2+28\lambda_8(\alpha_2+\lambda_4)\}\nonumber\\
&&+\frac{\lambda_{10}}{8\pi^2 (\alpha_0+\beta_0+m^4)^{1/2}},\\
%'P'O
\frac{\delta \lambda_{10}}{\delta t}&=& \frac{1}{128\pi^2(\alpha_0+\beta_0+m^4)^{9/2}}\{99225(\alpha_2+\lambda_4)^5\}\nonumber\\&&-\frac{1}{32\pi^2(\alpha_0+\beta_0+m^4)^{7/2}}\{23625(\alpha_2+\lambda_4)^3 (\alpha_4+\lambda_6)\}\nonumber\\
&&+\frac{1}{32\pi^2(\alpha_0+\beta_0+m^4)^{5/2}}\{4725(\alpha_2+\lambda_4)(\alpha_4+\lambda_6)^2+1890(\alpha_2+\lambda_4)^2\lambda_8\}\nonumber\\
&&-\frac{1}{16\pi^2(\alpha_0+\beta_0+m^4)^{3/2}} \{45\lambda_{10}(\alpha_2+\lambda_4)+210(\alpha_4+\lambda_6)\lambda_8\}, \\
%'RŽŸ
\frac{\delta \alpha_{2}}{\delta t}&=&\alpha_2-\frac{1}{48\pi^2(\alpha_0+\beta_0+m^4)^{7/2}}\{5(\alpha_0+2\beta_0)^2(\alpha_2+\lambda_4)^2\}\nonumber\\
&&+\frac{1}{32\pi^2(\alpha_0+\beta_0+m^4)^{5/2}}\{4\alpha_2(\alpha_0+2\beta_0)(\alpha_2+\lambda_4)+(3\alpha_0+10\beta_0)(\alpha_2+\lambda_4)^2\}\nonumber\\
&&-\frac{1}{48\pi^2(\alpha_0+\beta_0+m^4)^{3/2}}\{2\alpha_2^2+15\alpha_2(\alpha_2+\lambda_4)\}\nonumber\\
&&+\frac{\alpha_4}{8\pi^2(\alpha_0+\beta_0+m^4)^{1/2}},\\
%'TŽŸ
\frac{\delta \alpha_{4}}{\delta t}&=& \frac{1}{96\pi^2(\alpha_0+\beta_0+m^4)^{9/2}}\{175(\alpha_0+2\beta_0)^2(\alpha_2+\lambda_4)^3\}\nonumber\\
&&-\frac{1}{12\pi^2(\alpha_0+\beta_0+m^4)^{7/2}}\{30\alpha_2(\alpha_0+2\beta_0)(\alpha_2+\lambda_4)^2\nonumber\\
&&+5(3\alpha_0+10 \beta_0)(\alpha_2+\lambda_4)^3+5(\alpha_0+2\beta_0)^2(\alpha_2+\lambda_4)(\alpha_4+\lambda_6)\}\nonumber\\
&&+\frac{1}{32\pi^2(\alpha_0+\beta_0+m^4)^{5/2}}\{28\alpha_2^2(\alpha_2+\lambda_4)+93\alpha_2(\alpha_2+\lambda_4)^2\nonumber\\
&&+4(3\alpha_0+10\beta_0)(\alpha_2+\lambda_4)(\alpha_4+\lambda_6)
+8(\alpha_0+2\beta_0)(2\alpha_2\alpha_4+\alpha_4\lambda_4+\alpha_2\lambda_6)\}\nonumber\\
&&-\frac{1}{48\pi^2(\alpha_0+\beta_0+m^4)^{3/2}} (77\alpha_2\alpha_4+42\alpha_4\lambda_4+27\alpha_2\lambda_6)\label{eq2}.
\end{eqnarray}}

\begin{figure}{\begin{center}\includegraphics [width=70mm]{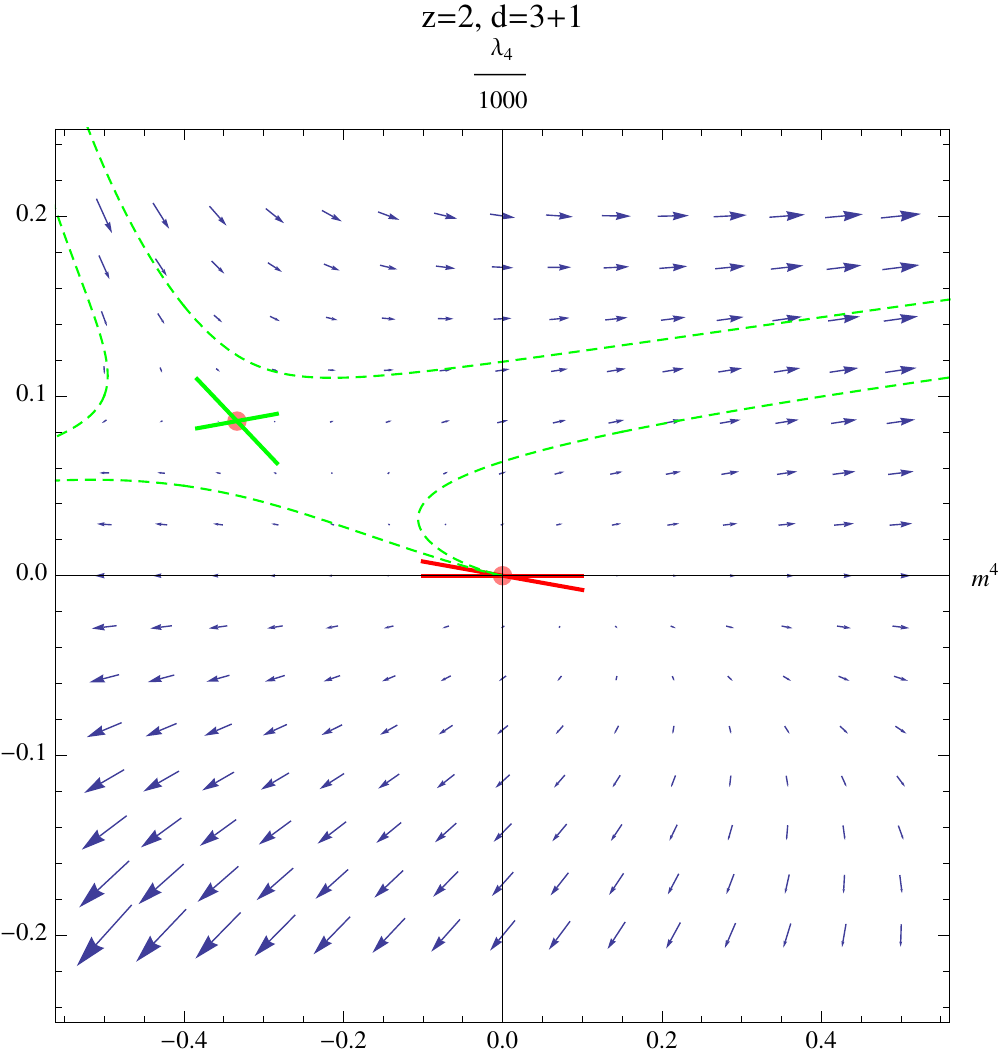}\end{center}}
\caption{Flow of $m^4, \lambda_4$ in  $z=2,~d=3+1$.}
\label{fig2}
\end{figure}

As an example, let us discuss the renormalization group flow in the theory subspace, in which only $m^4$ and $\lambda_4$ are nonzero. Equations~\eref{eq1000} and \eref{eq1001} then reduce to
\begin{eqnarray}
\frac{\delta m^4}{\delta t}&=&4m^4 +\frac{\lambda_4}{8\pi^2(1+m^4)^{1/2}},\\
\frac{\delta \lambda_4}{\delta t}&=&3\lambda_4 - \frac{3\lambda_4^2}{16\pi^2(1+m^4)^{3/2}},
\end{eqnarray}
where we take $\beta_0=1$ by rescaling the momentum. All other equations are satisfied trivially.  There are two fixed points given as
\begin{eqnarray}
&&m^4=0, \lambda_4=0, \nonumber\\
&&m^4=-\frac{1}{3}, \lambda_4=\frac{32}{3}\sqrt{\frac{2}{3}}\pi^2.
\end{eqnarray}
The first is a Gaussian fixed point, and the second is a nontrivial fixed point as seen in Fig.~\ref{fig2}. We would like to mention that this flow resembles the one in the ordinary scalar theory with the Lorentz symmetry in three space-time dimensions.

\subsection{Transformation of Variables} 
Our main interest is the restoration of the Lorentz symmetry in the IR region. To discuss the renormalization group flow in the IR region, it is useful to change the variables as  
\begin{eqnarray}
h=\frac{1}{\alpha_{0}},
\end{eqnarray}
and introduce new variables with hats~\cite{Anselmi:2008bq, Anselmi:2008bs, Dhar:2009dx}
\begin{eqnarray}
\hat{t}&=&h^{-\frac{1}{2}}t,\\
\hat{x}&=&x,\\
\hat{\phi}&=&h^{-\frac{1}{4}}\phi.
\end{eqnarray}
The action in the model \eref{eq0} with new variables is
\begin{eqnarray}
S &=& \int _ {\hat{t},  \hat{x}_i}    \Bigl\{
\frac{1}{2} ( \hat{\partial}_0 \hat{\phi}\hat{\partial}_0 \hat{\phi}+\hat{\partial}_i\hat{ \phi} \hat{\partial}_i\hat{ \phi}+\hat{m}^2\hat{\phi}^2 )\nonumber\\
&&+\frac{\hat{\lambda}_4}{4!} \hat{\phi}^4 +\frac{\hat{\lambda}_6}{6!} \hat{\phi}^6 +\frac{\hat{\lambda}_8}{8!} \hat{\phi}^8+\frac{\hat{\lambda}_{10} }{10!}  \hat{\phi}^{10}   \nonumber\\
&&+\frac{1}{2}h\hat{\partial}_i \hat{\partial}_i\hat{ \phi}\hat{ \partial}_j\hat{ \partial}_j\hat{ \phi} +\frac{1}{2} \Bigl(\frac{\hat{\alpha}_2}{2!} \hat{\phi}^2 \hat{\partial}_i \hat{\phi}\hat{ \partial}_i\hat{ \phi}+ \frac{\hat{\alpha}_4}{4!} \hat{\phi}^4 \hat{\partial}_i\hat{ \phi}\hat{ \partial}_i\hat{ \phi}\Bigl)\Bigl\},\label{eq4}
\end{eqnarray}
where
\begin{eqnarray}
&&\hat{m}^2\equiv h m^4
, \hat{\lambda}_4\equiv  h^{\frac{3}{2}}\lambda_4
,  \hat{\lambda}_6\equiv  h^{2}\lambda_6
,  \hat{\lambda}_8\equiv  h^{\frac{5}{2}} \lambda_8
,  \hat{\lambda}_{10}\equiv  h^{3} \lambda_{10},\nonumber\\
 &&\hat{\alpha}_2\equiv  h^{\frac{3}{2}}\alpha_2
, \hat{\alpha}_4\equiv  h^{2} \alpha_4,
\end{eqnarray}
where we also take $\beta_0=1$.  The dimensions in the unit of mass of new variables are as follows:
\begin{eqnarray}
&&[\hat{t}]=-1, [\hat{x}]=-1, [\hat{\phi}]=1, [\hat{m}^2]=2, [\hat{\lambda}_4]=0, [\hat{\lambda}_6]=-2,\nonumber\\
&&[\hat{\lambda}_8]=-4, [\hat{\lambda}_{10}]=-6, [h]=-2, [\hat{\alpha}_2]=-2, [\hat{\alpha}_4]=-4. 
\end{eqnarray}
They are identical to the canonical dimension in the Lorentz theory in four space-time dimensions. 

\begin{figure}{\begin{center}
\includegraphics [width=70mm]{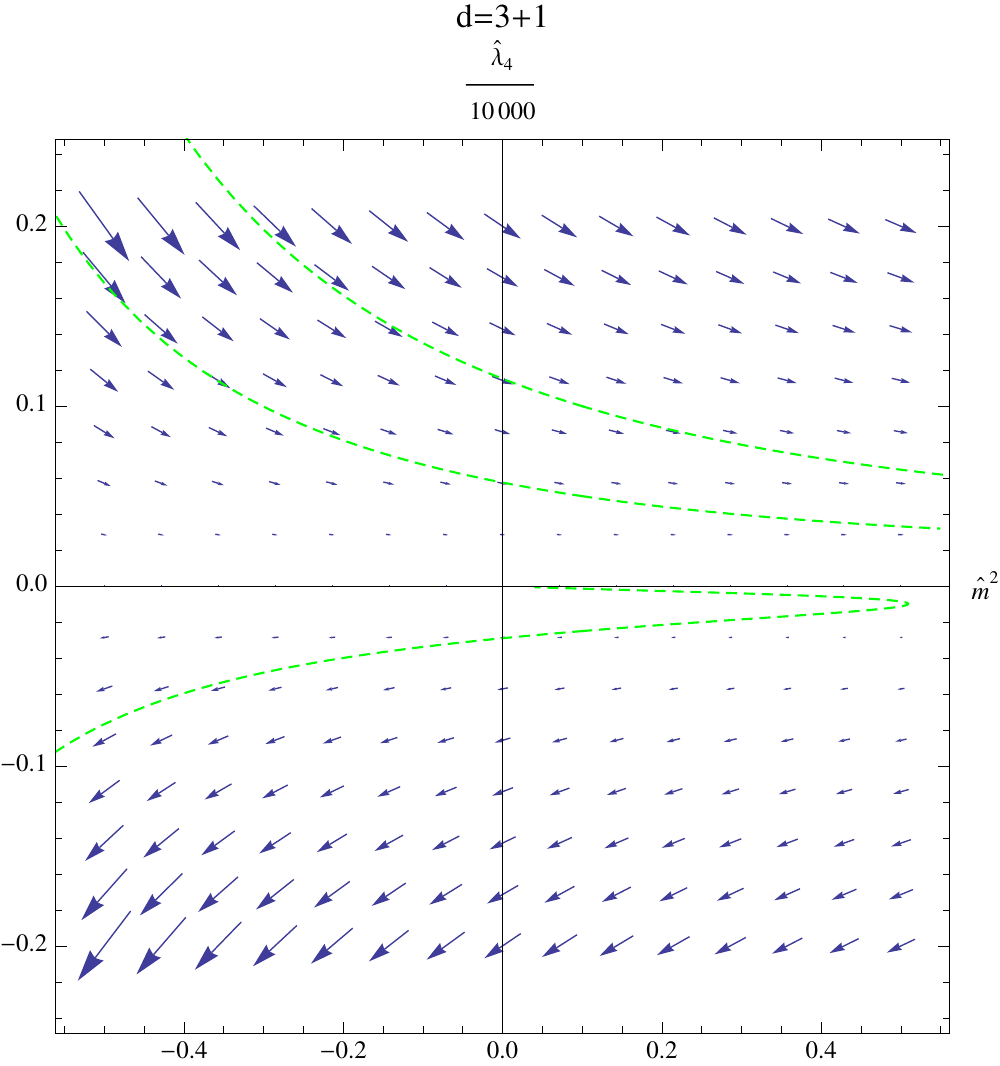}\end{center}}
\caption{Flow of $\hat{m}^2, \hat{\lambda}_4$ in $d=3+1$.}
\label{fig3}
\end{figure}

The renormalization group equations \eref{eq1}--\eref{eq2} in terms of the new variables (with hats) are given as \\
{\allowdisplaybreaks \begin{eqnarray}
\frac{\delta h}{\delta t}&=&-2h-\frac{\hat{\alpha_2} h}{8\pi^2(1+ h+\hat{m}^2 )^{1/2}},\label{eq2000}\\
\frac{\delta \hat{m}^2}{\delta t}&=&2\hat{m}^2 +\frac{1}{8\pi^2(1+ h+\hat{m}^2)^{1/2}}(\hat{\alpha}_2+\hat{\lambda}_4-\hat{\alpha}_2\hat{m}^2),\\
\frac{\delta \hat{\lambda}_4}{\delta t}&=&- \frac{1}{16\pi^2(1+ h+\hat{m}^2 )^{3/2}}\{3(\hat{\alpha}_2+\hat{\lambda}_4)^2\}\nonumber\\
&&+\frac{1}{16\pi^2(1+h+\hat{m}^2 )^{1/2}}\{2(\hat{\alpha}_4+\hat{\lambda}_6)-3\hat{\lambda}_4\hat{\alpha}_2\},  \\
\frac{\delta \hat{\lambda}_6}{\delta t}&=&-2\hat{\lambda}_6 + \frac{1}{32\pi^2(1+h+\hat{m}^2 )^{5/2}}\{45(\hat{\alpha}_2+\hat{\lambda}_4)^3\}\nonumber\\
&&- \frac{1}{16\pi^2(1+h+\hat{m}^2 )^{3/2}}\{15(\hat{\alpha}_2+\hat{\lambda}_4)(\hat{\alpha}_4+\hat{\lambda}_6)\}\nonumber\\
&&+\frac{1}{8\pi^2 (1+ h+\hat{m}^2 )^{1/2}}(\hat{\lambda}_8-2\hat{\lambda}_6\hat{\alpha}_2),  \\
\frac{\delta \hat{\lambda}_{8}}{\delta t}&=&-4{\hat{\lambda}}_8 -\frac{1}{64\pi^2(1+ h+\hat{m}^2 )^{7/2}}\{1575(\hat{\alpha}_2+\hat{\lambda}_4)^4\}\nonumber\\
&&+\frac{1}{16\pi^2(1+ h+\hat{m}^2)^{5/2}}\{315(\hat{\alpha}_2+\hat{\lambda}_4)^2(\hat{\alpha}_4+\hat{\lambda}_6) \}\nonumber\\
&&-\frac{1}{16\pi^2 (1+ h+\hat{m}^2 )^{3/2}}\{35(\hat{\alpha}_4+\hat{\lambda}_6)^2+28\hat{\lambda}_8(\hat{\alpha}_2+\hat{\lambda}_4)\}\nonumber\\
&&+\frac{1}{16\pi^2 (1+ h+\hat{m}^2)^{1/2}}(2\hat{\lambda}_{10}-5\hat{\alpha}_2\hat{\lambda}_8),  \\%ok
%'P'O
\frac{\delta\hat{ \lambda}_{10}}{\delta t}&=&-6\hat{\lambda}_{10}+ \frac{1}{128\pi^2(1+ h+\hat{m}^2 )^{9/2}}\{99225(\hat{\alpha}_2+\hat{\lambda}_4)^5\}\nonumber\\
&&-\frac{1}{32\pi^2(1+h+\hat{m}^2 )^{7/2}}\{23625(\hat{\alpha}_2+\hat{\lambda}_4)^3 (\hat{\alpha}_4+\hat{\lambda}_6)\}\nonumber\\
&&+\frac{1}{32\pi^2(1+h+\hat{m}^2 )^{5/2}}\{4725(\hat{\alpha}_2+\hat{\lambda}_4)(\hat{\alpha}_4+\hat{\lambda}_6)^2+1890(\hat{\alpha}_2+\hat{\lambda}_4)^2\hat{\lambda}_8\}\nonumber\\
&&-\frac{1}{16\pi^2(1+h+\hat{m}^2 )^{3/2}}\{ 45\hat{\lambda}_{10}(\hat{\alpha}_2+\hat{\lambda}_4)+210(\hat{\alpha}_4+\hat{\lambda}_6)\hat{\lambda}_8\} \nonumber\\
&&-\frac{ 3\hat{\lambda}_{10}\hat{\alpha}_2}{8\pi^2(1+h+\hat{m}^2 )^{1/2}},\\   %ok
%'R
\frac{\delta \hat{\alpha}_{2}}{\delta t}&=&-2\hat{\alpha}_2-\frac{1}{48\pi^2(1+h+\hat{m}^2 )^{7/2}}\{5(1+2 h)^2(\hat{\alpha}_2+\hat{\lambda}_4)^2\}\nonumber\\
&&+\frac{1}{32\pi^2(1+h+\hat{m}^2 )^{5/2}}\{4\hat{\alpha}_2(1+2h)(\hat{\alpha}_2+\hat{\lambda}_4)+(3+10h)(\hat{\alpha}_2+\hat{\lambda}_4)^2\}\nonumber\\
&&-\frac{1}{48\pi^2(1+ h+\hat{m}^2 )^{3/2}}\{2\hat{\alpha}_2^2+15\hat{\alpha}_2(\hat{\alpha}_2+\hat{\lambda}_4)\}\nonumber\\
&&+\frac{1}{16\pi^2(1+ h+\hat{m}^2)^{1/2}}(2\hat{\alpha}_4 -3\hat{\alpha}_2^2),\\%ok
%'T
\frac{\delta \hat{\alpha}_{4}}{\delta t}&=& -4\hat{\alpha}_4+\frac{1}{96\pi^2(1+ h+\hat{m}^2 )^{9/2}}\{175(1 +2h)^2(\hat{\alpha}_2+\hat{\lambda}_4)^3\}\nonumber\\
&&-\frac{1}{12\pi^2(1+ h+\hat{m}^2 )^{7/2}}\{30\hat{\alpha}_2(1+2h)(\hat{\alpha}_2+\hat{\lambda}_4)^2+5(3+10  h)(\hat{\alpha}_2+\hat{\lambda}_4)^3\nonumber\\
&&+5(1+2 h)^2(\hat{\alpha}_2+\hat{\lambda}_4)(\hat{\alpha}_4+\hat{\lambda}_6)\}\nonumber\\
&&+\frac{1}{32\pi^2(1+h+\hat{m}^2 )^{5/2}}\{28\hat{\alpha}_ {2}^2(\hat{\alpha}_2+\hat{\lambda}_4)+93\hat{\alpha}_2(\hat{\alpha}_2+\hat{\lambda}_4)^2\nonumber\\
&&+4(3+10 h)(\hat{\alpha}_2+\hat{\lambda}_4)(\hat{\alpha}_4+\hat{\lambda}_6)+8(1+2 h)(2\hat{\alpha}_2\hat{\alpha}_4+\hat{\alpha}_4\hat{\lambda}_4+\hat{\alpha}_2\hat{\lambda}_6)\}\nonumber\\
&&-\frac{1}{48\pi^2(1+ h+\hat{m}^2 )^{3/2}}(77\hat{\alpha}_2\hat{\alpha}_4+42\hat{\alpha}_4\hat{\lambda}_4+27\hat{\alpha}_2\hat{\lambda}_6)\nonumber\\
&& -\frac{1}{4\pi^2(1+ h+\hat{m}^2 )^{1/2}} (\hat{\alpha}_2\hat{\alpha}_4).\label{eq2001}
\end{eqnarray}}
If we set $h=0, \hat{\alpha}_2=0,$ and $\hat{\alpha}_4=0$, these renormalization group equations exactly coincide with the equations in the case of the local potential approximation in the ordinary theory, which has Lorentz symmetry, as expected. As an example, we give the renormalization group flow in the theory subspace, in which only $\hat{m}^2$ and $\hat{\lambda}_4$ are nonzero. (See Fig.~\ref{fig3}.) 

These renormalization group equations have a Gaussian fixed point at 
\begin{eqnarray}
&&\hat{m}^2=0, \hat{\lambda}_4=0, \hat{\lambda}_6=0, \hat{\lambda}_8=0, \hat{\lambda}_{10}=0, h=0, \hat{\alpha}_2=0, \hat{\alpha}_4=0.\label{eq3000}\hspace{1cm}
\end{eqnarray}
In the neighborhood of the fixed point, the renormalization group equations \eref{eq2000}--\eref{eq2001} can be approximated as{\allowdisplaybreaks 
\begin{eqnarray}
\frac{\delta h}{\delta t}&=&-2h,\label{eq4000}\\
\frac{\delta \hat{m}^2}{\delta t}&=&2\hat{m}^2 +\frac{\hat{\alpha}_2+\hat{\lambda}_4}{8\pi^2},\\
\frac{\delta \hat{\lambda}_4}{\delta t}&=&\frac{\hat{\alpha}_4+\hat{\lambda}_6 }{8\pi^2},  \\
\frac{\delta \hat{\lambda}_6}{\delta t}&=&-2\hat{\lambda}_6 +\frac{\hat{\lambda}_8}{8\pi^2 },  \\%
\frac{\delta \hat{\lambda}_{8}}{\delta t}&=&-4{\hat{\lambda}}_8 +\frac{\hat{\lambda}_{10}}{8\pi^2 },  \\%ok
%'P'O
\frac{\delta\hat{ \lambda}_{10}}{\delta t}&=&-6\hat{\lambda}_{10},\\
%'R
\frac{\delta \hat{\alpha}_{2}}{\delta t}&=&-2\hat{\alpha}_2+\frac{\hat{\alpha}_4}{8\pi^2},\\%ok
%'T
\frac{\delta \hat{\alpha}_{4}}{\delta t}&=& -4\hat{\alpha}_4\label{eq4001},
\end{eqnarray}}at the linear order in the perturbation theory. From Eq.~\eref{eq4000}, it turns out that the fixed point \eref{eq3000} is the IR one. These equations \eref{eq4000}--\eref{eq4001} tell us that when we increase the energy scale, $\hat{m}^2$ and $\hat{\lambda}_4$ become dominant compared with $h$, $\hat{\alpha}_{2}$, and $\hat{\alpha}_{4}$, which are the coupling constants of terms breaking the Lorentz symmetry. This implies the restoration of the Lorentz symmetry in the neighborhood of the Gaussian fixed point in the IR region. They coincide with the expectation by power counting.  However, the purpose of this paper is to study the restoration of the Lorentz symmetry nonperturbatively. Thus, we should solve Eqs.~\eref{eq1}--\eref{eq2} without linear approximations.  In the next section, we perform these calculations by numerical analysis.

\section{Numerical Analysis}\label{NA}
The terms in the third line in Eq.~\eref{eq4} violate the Lorentz symmetry.  If they vanish in the IR region, we can say that the Lorentz symmetry is restored in the IR region. In the following, solving the Wegner-Houghton equations \eref{eq1}--\eref{eq2} with some initial conditions by numerical analysis, we study the renormalization group flow of the Lifshitz-type theory to see if this is the case. 

We should choose the initial conditions carefully. To obtain flows of proper physical theories, we looked for initial conditions that satisfy two requirements. First, the coupling constants that violate the Lorentz symmetry should become negligible at low energy. Second, all the coupling constants should approach the Lifshitz fixed point at high energy to obtain UV complete theories. The key problem here is whether such initial conditions exist. Actually, we found flows with typical initial conditions that satisfy the two requirements. In this paper, as examples, we give results for two initial conditions (case 1 and 2) defined as follows:\\
\noindent Case 1: The initial conditions are\begin{eqnarray}&&m^4=1.00\times10^{-4}, \lambda_4=8.50\times10^{-2}, \lambda_6=5.00\times10^{-1}, \lambda_8=5.00\times10^{-1}, \nonumber\\
&&\lambda_{10}=0, \alpha_0=7.50\times10^{-1}, \alpha_2=3.50\times10^{-1}, \alpha_4=0.\nonumber
\end{eqnarray}
In terms of hatted coupling constants, they are
\begin{eqnarray}
&&\hat{m}^2=1.33\times10^{-4}, \hat{\lambda}_4=1.31\times10^{-1}, \hat{\lambda}_6=8.89\times10^{-1}, \hat{\lambda}_8=1.03, \nonumber\\
&&\hat{\lambda}_{10}=0, h=1.33, \hat{\alpha}_2=5.39\times10^{-1}, \hat{\alpha}_4=0,\nonumber
\end{eqnarray}
which are written to three significant figures.

\begin{figure}[!t]{\begin{center}\includegraphics [width=78mm]{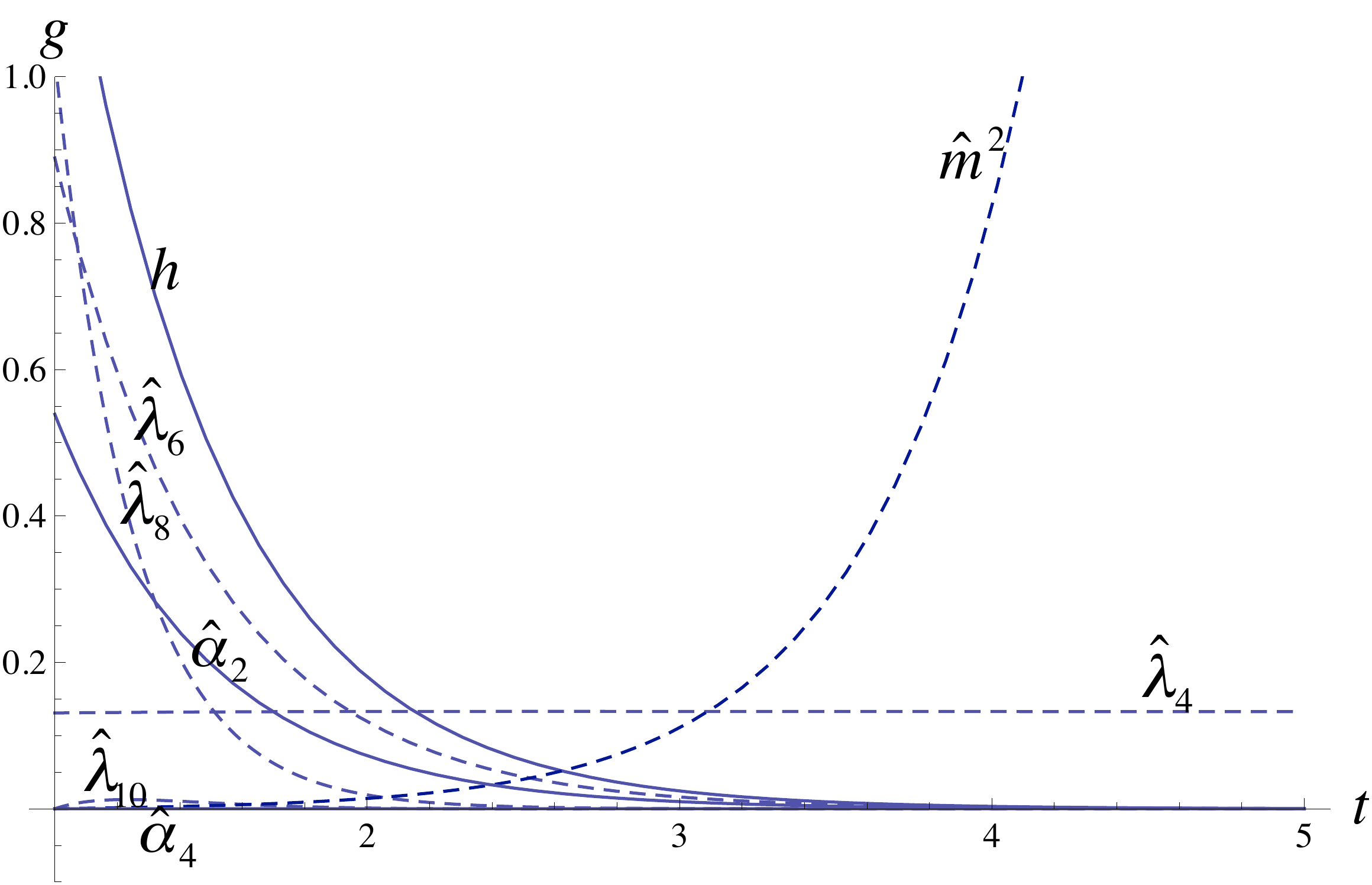}\end{center}}
\caption{Renormalization group flow of hatted coupling constants against decreasing energy scale $t$ under the initial condition in case 1. The continuous lines show the flows of the coupling constants $h, \hat{\alpha}_2$, and $\hat{\alpha}_4$, which violate the Lorentz symmetry. The dashed lines show the ones of $\hat{m}^2, \hat{\lambda}_4, \hat{\lambda}_6, \hat{\lambda}_8$, and $\hat{\lambda}_{10}$.}
\label{fig4}
\end{figure}

\begin{figure}[!t]{\begin{center}\includegraphics [width=78mm]{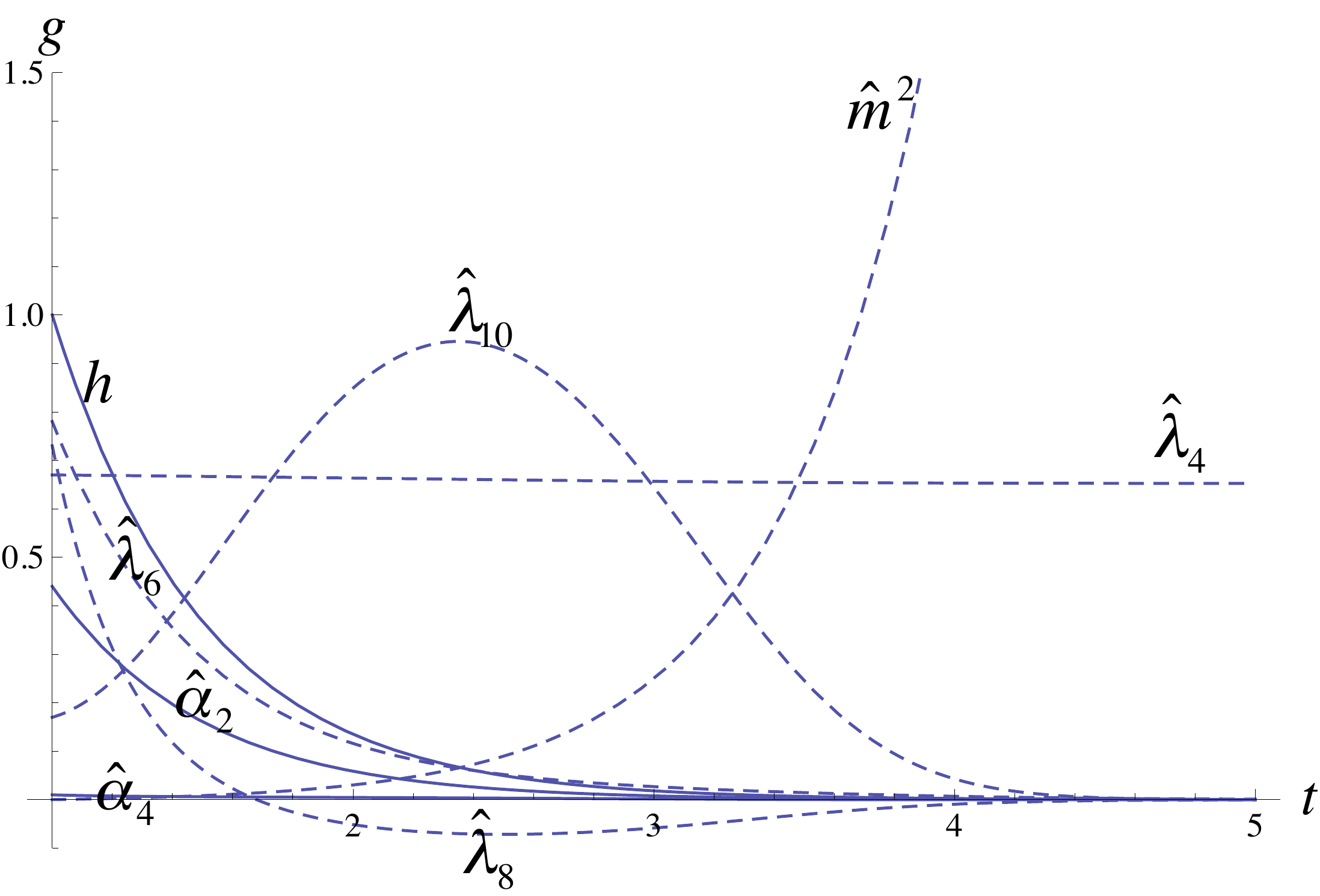}\end{center}}
\caption{Renormalization group flow of hatted coupling constants against decreasing energy scale $t$ under the initial condition in case 2.}
\label{fig5}
\end{figure}

\noindent Case 2: The initial conditions are
\begin{eqnarray}&&m^4=1.00\times10^{-4}, \lambda_4=6.70\times10^{-1}, \lambda_6=7.80\times10^{-1}, \lambda_8=7.30\times10^{-1}, \nonumber\\
&&\lambda_{10}=1.70\times10^{-1}, \alpha_0=1.00, \alpha_2=4.40\times10^{-1}, \alpha_4=1.00\times10^{-2}.\nonumber
\end{eqnarray}
In terms of hatted coupling constants, they are
\begin{eqnarray}
&&\hat{m}^2=1.00\times10^{-4}, \hat{\lambda}_4=6.70\times10^{-1}, \hat{\lambda}_6=7.80\times10^{-1}, \hat{\lambda}_8=7.30\times10^{-1}, \nonumber\\
&&\hat{\lambda}_{10}=1.70\times10^{-1}, h=1.00, \hat{\alpha}_2=4.40\times10^{-1}, \hat{\alpha}_4=1.00\times10^{-2}.\nonumber
\end{eqnarray}
\noindent Figures~\ref{fig4} and \ref{fig5} show the renormalization group flows of the coupling constants with hatted variables for cases 1 and 2, respectively. The results show that $h, \hat{\alpha}_2$, and $\hat{\alpha}_4$ rapidly become negligible with decreasing energy scale $(t > 0)$; therefore, the third line terms of Eq.~\eref{eq4} turn out to be highly suppressed. This implies that the Lorentz symmetry is restored in the IR region.

\begin{figure}[t]{\begin{center}\includegraphics [width=78mm]{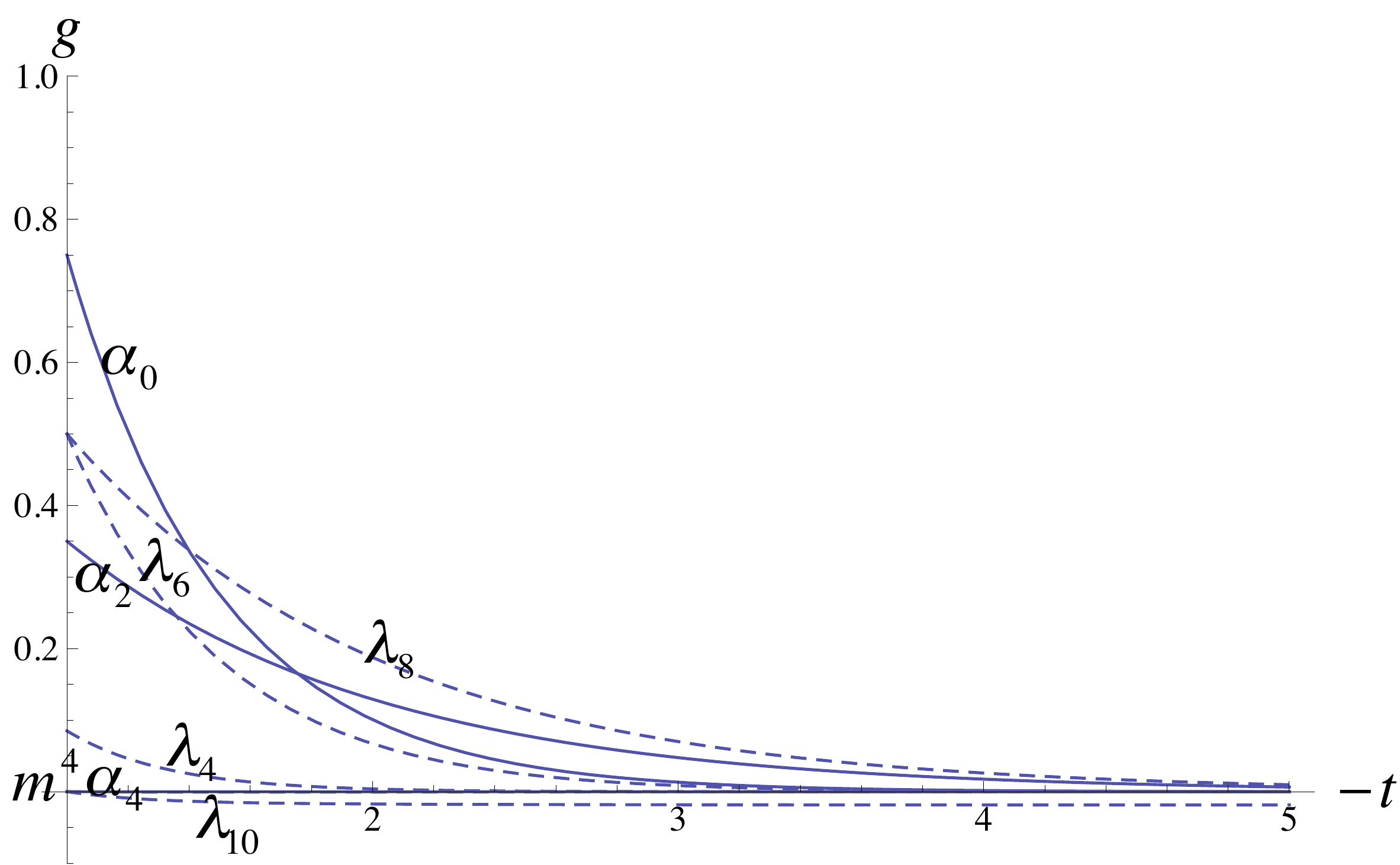}\end{center}}
\caption{Renormalization group flow of unhatted coupling constants against increasing energy scale $-t$ under the initial condition in case 1. The continuous lines show the flows of the coupling constants $\alpha_0, {\alpha}_2$, and ${\alpha}_4$. The dashed lines show the ones of ${m}^4, {\lambda}_4, {\lambda}_6, {\lambda}_8$, and ${\lambda}_{10}$.}
\label{fig6}
\end{figure}

\begin{figure}[!t]{\begin{center}\includegraphics [width=78mm]{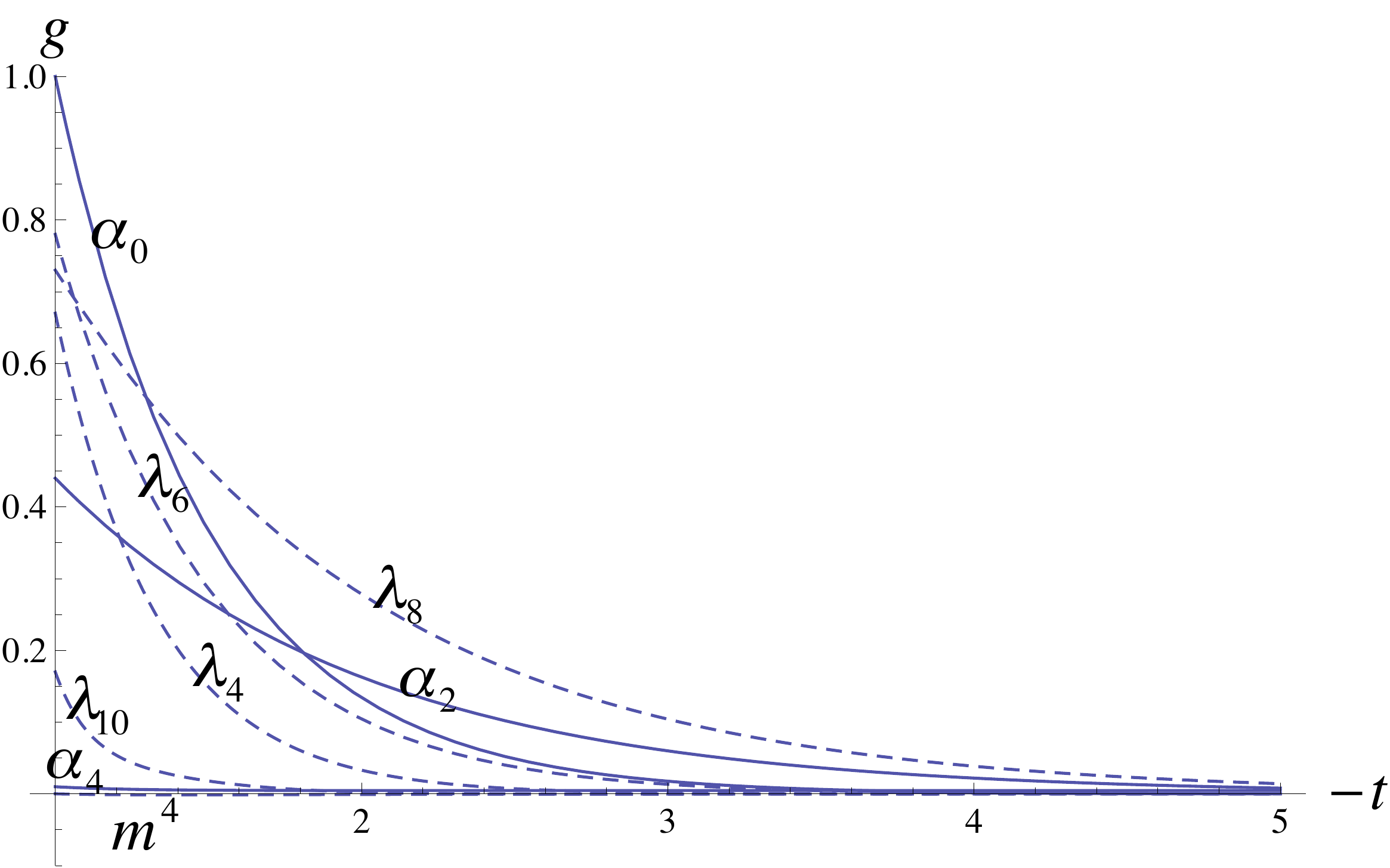}\end{center}}
\caption{Renormalization group flow of unhatted coupling constants against increasing energy scale $-t$ under the initial condition in case 2.}
\label{fig7}
\end{figure}

On the other hand, Figs.~\ref{fig6} and \ref{fig7} show the renormalization group flows of the unhatted coupling constants with increasing energy scale $(t < 0)$ for cases 1 and 2. The results show that all the coupling constants approach the Lifshitz fixed point with increasing energy. Therefore, we obtain the UV complete renormalizable theories, which have the Lorentz symmetry in the IR region, under proper initial conditions. 


Furthermore, case 2 is very interesting. Remarkably, the theory in case 2 has nonzero coupling constants of the interaction term, $\hat{\lambda}_4, \hat{\lambda}_6, \hat{\lambda}_8, \hat{\lambda}_{10}$, even when the Lorentz symmetry is restored at low energy. For example, at the energy scale $t=3.5$ in Fig.~\ref{fig5}, the coupling constants are
\begin{eqnarray}
&&\hat{m}^2=6.85\times10^{-1}, \hat{\lambda}_4=6.55\times10^{-1}, \hat{\lambda}_6=1.46\times10^{-2}, \hat{\lambda}_8=-3.06\times10^{-2}, \nonumber\\
&&\hat{\lambda}_{10}=2.46\times10^{-1}, h=6.72\times10^{-3}, \hat{\alpha}_2=3.01\times10^{-3}, \hat{\alpha}_4=2.45\times10^{-4},\nonumber
\end{eqnarray}
which are written to three significant figures.

 Finally, we would like to mention a possibility for other initial conditions. Certainly, it is possible that there are other interesting initial conditions, and to classify the regions of the flows generally is an interesting future work. The most important thing, however, is that there exists at least one such flow.

\section{Summary and Discussion}\label{SD}

In the Lifshitz-type theory, higher derivative terms in the kinetic terms suppress the UV divergence. However, there is a problem of broken Lorentz symmetry; therefore, we should examine whether the theory restores the Lorentz symmetry in the IR region.  In this paper, we applied the Wegner-Houghton equation with the momentum cutoff in cylindrical shape and analyzed the $z=2, d=3+1$ Lifshitz-type scalar model numerically.  We find that the terms that break the Lorentz symmetry vanish at low energy. Remarkably, the Lifshitz-type theory has nontrivial interaction terms $\hat{\lambda}_{n}\phi^{n} (n=4, 6, 8, 10)$ even when the Lorentz symmetry is restored at low energy. We find a concrete solution to the long-standing problem of triviality of $\phi^4$ theory in $d=3+1$. In summary, $z=2, d=3+1$ Lifshitz-type scalar theory, at least for the model in this paper, restores the Lorentz symmetry in the IR region, and we obtain a UV complete renormalizable theory under proper initial conditions.

It would also be interesting to analyze multiple fields including fermions.  They are also soluble by this method in principle, although some improvements may be needed in this analysis.  The truncation method remains as a matter to be discussed further. Given some symmetries, we may improve this analytic method. There is also room for discussion on the cutoff method.  Nonetheless, our method is attractive in the sense that we can judge one of the consistencies of the Lifshitz-type theories.  Further studies using this method will be needed. 
%%%%%%%%%%%%%%%%%%%%%%%%%%%%%%%%%%%
\section*{Acknowledgments}
We would like to thank Kiyoshi Higashijima and Etsuko Itou for very stimulating discussions and  meaningful advice throughout this work.  We also thank Tetsuya Onogi and Satoshi Yamaguchi for many useful comments.
%%%%%%%%%%%%%%%%%%%%%%%%%%%%%%%%%%5

\appendix
\section{Notation}\label{appa}
In this paper, we use the following notation.
The definitions of integral symbols are
\begin{eqnarray}
\int_x&\equiv& \int d^{d}x,\\
\int_{p}&\equiv& \int\frac{d^{d}p}{(2\pi)^d}.
\end{eqnarray}
The definitions of $\delta$ functions and its symbols are
\begin{eqnarray}
\int_{x}\delta(x)g(x)&=&g(0),\\
\int_{p}\hat{\delta}(p)f(p)&=&f(0),
\end{eqnarray}
\begin{eqnarray}
\delta(x)&=&\int_{p}e^{-ipx},\\
\hat{\delta}(p)&\equiv&(2\pi )^d\delta(p)=\int_{x}e^{ipx}.
\end{eqnarray}
The definitions of Fourier transformation of fields are
\begin{eqnarray}
\phi(x)&=&\int_{p}e^{ipx}\phi(p),\\
\phi(p)&=&\int_{x}e^{-ipx}\phi(x).
\end{eqnarray}
The definition of trace symbol is
\begin{eqnarray}
\mathrm{tr}\equiv\int_k\int_{k'}\hat{\delta}(k-k').
\end{eqnarray}

\section{Derivation of Wegner-Houghton Equation}\label{appb}

{\def\thesection{\Alph{section}}
This is a review of \cite{Clark:1992jr, Aoki:2000wm}.  The general Wilsonian effective action is given as 
\begin{eqnarray}
S[\Omega;\Lambda]&=&\sum_n\frac{1}{n!}\int_{p_1}\cdots\int_{p_n}\hat{\delta}^{(D)}(p_1+\cdots+p_n)g_{{i_1},...,{i_n}}(p_1,\cdots,p_n;\Lambda)\nonumber\\
&&\hspace{2.5cm}\times \Omega_{i_1}(p_1;\Lambda)\cdots\Omega_{i_n}(p_n;\Lambda),
\end{eqnarray}
where 
\begin{eqnarray}
\Omega(p_i;\Lambda)&\equiv&\Omega(p_i)\theta(\Lambda-p_i).
\end{eqnarray}
$\Omega(p_i)$ denotes general fields, for example, in the case of scalar fields $\Omega(p_i)=\phi(p_i)$.  We introduce the shell-mode wave functions $\Omega_{s}(p_i)$, which are nonzero only for $\Lambda(\delta t)=\Lambda e^{-\delta t}\leq p_i \leq \Lambda,$ and  $\Omega_{IR}(p_i;\Lambda(\delta t))$, which are nonzero only for $p_i \leq \Lambda(\delta t)$, where $\Lambda(t)\equiv \Lambda e^{-t}, \delta \Lambda \equiv \Lambda \delta t$. We then write
\begin{eqnarray}
\Omega(p_i;\Lambda)&=&\Omega_{IR}(p_i;\Lambda(\delta t))+\Omega_{s}(p_i)\label{eq5000}.
\end{eqnarray}
The partition function $Z$ is given as
\begin{eqnarray}
Z&=&\int [d\Omega]\exp\{ {-S[ \Omega; \Lambda]} \}.
\end{eqnarray}
Using Eq.~\eref{eq5000}, we split $[d\Omega]$ into $[d\Omega_{IR}]$ and $[d\Omega_{s}]$ in the partition function $Z$ to obtain
\begin{eqnarray}
Z&=&\int [d\Omega_{IR}] \int [d\Omega_s]\exp\{ {-S[ \Omega_{IR}+\Omega_s;\Lambda ]} \}.\label{eq5001}
\end{eqnarray}
On the other hand, a partition function that is defined by the cutoff $\Lambda(\delta t)$ is given as 
\begin{eqnarray}
Z&=&\int [d\Omega_{IR}]\exp\{ {-S[ \Omega_{IR}; \Lambda(\delta t)]} \},
\end{eqnarray}
which gives the same value as \eref{eq5001}. Therefore, we obtain the renormalization group equation
\begin{eqnarray}
\exp\{ {-S[ \Omega_{IR}; \Lambda(\delta t)]} \} = \int [d\Omega_s]\exp\{ {-S[ \Omega_{IR}+\Omega_s;\Lambda ]} \}.\label{eq5002}
\end{eqnarray}
On the LHS of Eq.~\eref{eq5002},  expanding $S[ \Omega_{IR}+\Omega_s;\Lambda ]$ by $\Omega_s$ and integrating out $\Omega_s$ through the first order of $\delta \Lambda$, we obtain the equation
\begin{eqnarray}
S[\Omega_{IR};\Lambda(\delta t)]&=&S[\Omega_{IR};\Lambda]+\frac{1}{2}\mathrm{tr} \ln \Bigl(\frac{\delta^2 S}{\delta\Omega^i \delta\Omega^j }\Bigl)-\frac{1}{2}\frac{\delta S}{\delta \Omega^i }\Bigl(\frac{\delta^2 S}{\delta\Omega^i \delta\Omega^j }\Bigl)^{-1}\frac{\delta S}{\delta \Omega^j}.\hspace{1cm}\label{eq21}
\end{eqnarray}

On the other hand, a general action at the scale $\Lambda(\delta t)$ or $\Lambda$ is 
\begin{eqnarray}
S[\Omega_{IR};\Lambda(\delta t)]&=&\sum_n\frac{1}{n!}\int_{p_1}^{\Lambda(\delta t)}\cdots\int_{p_n}^{\Lambda(\delta t)}\hat{\delta}^{(D)}(p_1+\cdots+p_n)g_{n}(p; \Lambda(\delta t))\nonumber\\
&&\times\Omega_{i_1}(p_1;\Lambda(\delta t))\cdots\Omega_{i_n}(p_n;\Lambda(\delta t)),\\
S[\Omega_{IR};\Lambda]&=&\sum_n\frac{1}{n!}\int_{p_1}^{\Lambda(\delta t)}\cdots\int_{p_n}^{\Lambda(\delta t)}\hat{\delta}^{(D)}(p_1+\cdots+p_n)g_{n}(p; \Lambda)\nonumber\\
&&\times\Omega_{i_1}(p_1;\Lambda(\delta t))\cdots\Omega_{i_n}(p_n;\Lambda(\delta t)),
\end{eqnarray}
where $p\equiv \{p_1, p_2, ... , p_n\}$. The difference in effective actions is given by
\begin{eqnarray}
S[\Omega_{IR};\Lambda(\delta t)]-S[\Omega_{IR};\Lambda]\hspace{7cm}\nonumber\\
=\sum_n\frac{1}{n!}\int_{p_1}^{\Lambda(\delta t)}\cdots\int_{p_n}^{\Lambda(\delta t)}\hat{\delta}^{(D)}(p_1+\cdots+p_n)[g_{n}(p; \Lambda(\delta t))-g_{n}(p; \Lambda)]\nonumber\\
\times\Omega_{i_1}(p_1;\Lambda(\delta t))\cdots\Omega_{i_n}(p_n;\Lambda(\delta t))\hspace{1cm}\nonumber\\
=\sum_n \frac{1}{n!} \int_{p_1}^{\Lambda(\delta t)}\cdots\int_{p_n}^{\Lambda(\delta t)}\hat{\delta}^{(D)}(p_1+\cdots+p_n)\Bigl[-(\delta t)\Lambda \frac{\partial g_{n}(p; \Lambda)}{\partial \Lambda}\Bigl]\hspace{0.8cm}\nonumber\\
\times\Omega_{i_1}(p_1;\Lambda(\delta t))\cdots\Omega_{i_n}(p_n;\Lambda(\delta t)).\hspace{1cm}\label{eq000}
\end{eqnarray}
As mentioned above, we can write the difference in terms of the coupling constant differentiated with respect to $\Lambda$.  We
 define $p=\Lambda\hat{p}$.  When this coupling depends on the momentum explicitly, it is written as 
\begin{eqnarray}
\delta g(\Lambda\hat{p}; \Lambda)&=&\frac{\partial g(p; \Lambda)}{\partial\Lambda}\delta\Lambda+\sum_{i=1}^{n}\frac{\partial g(\Lambda\hat{p}; \Lambda)}{\partial\hat{p}_i}\delta \hat{p}_i\nonumber\\
&=&-\frac{\partial g(p; \Lambda)}{\partial\Lambda}\Lambda\delta t+\sum_{i=1}^{n}\frac{\partial g(\Lambda\hat{p}; \Lambda)}{\partial\hat{p}_i}\hat{p}_i\delta t.
\end{eqnarray}
Therefore, we obtain
\begin{eqnarray}
\Lambda\frac{\partial}{\partial\Lambda}g(p; \Lambda)=\Lambda \frac{d}{d\Lambda}g(p; \Lambda)-\sum_{i=1}^{n} \hat{p}_i \frac{\partial}{\partial \hat{p}_i}g(p; \Lambda).
\end{eqnarray}
The dimensions of fields become $d_\Omega-\gamma$ as a consequence of the quantum effect.  Using a dimensionless coupling $\hat{g}$, we write $g(\Lambda)=\Lambda^{[g]}\hat{g}(\Lambda)$, where $[g](=d-\sum_{\Omega_i}(d_{\Omega_i}-\gamma))$ is the dimensions of $g$; then
\begin{eqnarray}
\Lambda\frac{\partial}{\partial\Lambda}g(\Lambda)&=&\Lambda^{[g]}\Lambda \frac{d}{d\Lambda}\hat{g}(\Lambda)+[g]g(\Lambda)-\sum_{i=1}^{n} \hat{p}_i \frac{\partial}{\partial \hat{p}_i}g(\Lambda)\nonumber\\
&=&\Lambda^{[g]}\Lambda \frac{d}{d\Lambda}\hat{g}(\Lambda)+\Bigl\{d-\sum_{\Omega_i}(d_{\Omega_i}-\gamma)\Bigl\}g(\Lambda)-\sum_{i=1}^{n} \hat{p}_{i}\frac{\partial}{\partial \hat{p}_{i}}g(\Lambda).\hspace{1cm}\label{eq001}
\end{eqnarray}
Substituting \eref{eq001} in \eref{eq000}, and writing them in terms of the action $S$, we obtain
\begin{eqnarray}
S[\Omega_{IR};\Lambda(\delta t)]-S[\Omega_{IR};\Lambda]\hspace{6cm}\nonumber\\
=\delta t \Bigl\{ -\Lambda \frac{d}{d\Lambda} S-dS+\int_{p}\Omega_{p}^{i}\Bigl(d_{\Omega}-\gamma+{\hat{p}^{\mu}} \frac{\partial'}{\partial {\hat{p}^{\mu}}} \Bigl) \frac{\delta}{\delta \Omega_{p}^{i}}S\Bigl\} \label{eq22}.
\end{eqnarray}
The term $\Lambda \frac{d}{d\Lambda} S$ denotes $\sum_n\frac{1}{n!}\int_{p_1}\cdots\int_{p_n}\hat{\delta}^{(D)}(p_1+\cdots+p_n)\Lambda\frac{d}{d\Lambda}\hat{g}_{{i_1},...,{i_n}} \hat{\Omega}_{i_1}\cdots\hat{\Omega}_{i_n}$, where $\hat{\Omega}$ is a dimensionless field. The operator $\int_{p}\Omega_{p}^{i}(d_{\Omega}-\gamma)\frac{\delta}{\delta \Omega_{p}^{i}}$ of the third term on the RHS counts the degrees of powers of the fields multiplied by $(d_{\Omega}-\gamma)$ for each term.  The operator $\int_{p}\Omega_{p}^{i}{\hat{p}^{\mu}} \frac{\partial'}{\partial {\hat{p}^{\mu}}} \frac{\delta}{\delta \Omega_{p}^{i}}$ of the third term on the RHS counts the number of derivative operators for each term.  The prime mark of $\partial'$ denotes that the operator does not operate on momentum arguments of delta functions or step functions but only on the coupling constants.  Comparing Eq.~\eref{eq21} with Eq.~\eref{eq22}, we finally obtain the Wegner-Houghton equation \eref{eq00}.}

\section{Main Calculations}\label{appc}

We perform integration by parts and Fourier transformation in the action \eref{eq0} to obtain 
\begin{eqnarray}
S &=& \int _ {q_1 q_2} \frac{1}{2}    (q_{1,0}^2+\alpha_0 q_{1,i}^2+\beta_0 q_{1,i}^2q_{2,j}^2+m^4)\phi_{q_1}\phi_{q_2}\hat{\delta}(q_1+q_2)\nonumber\\
&&+\int _ {q_1 \cdots q_4}\Bigl(\frac{1}{12}\alpha_2 q_{1,i}^2+\frac{\lambda_4}{4!}\Bigl)\phi_{q_1}\cdots \phi_{q_4}\hat{\delta}(q_1+\cdots+q_4)\nonumber\\
&&+\int _ {q_1 \cdots q_6}\Bigl(\frac{1}{240}\alpha_4 q_{1,i}^2+\frac{\lambda_6}{6!}\Bigl)\phi_{q_1}\cdots \phi_{q_6}\hat{\delta}(q_1+\cdots+q_6)\nonumber\\
&&+\int _ {q_1 \cdots q_8}\frac{\lambda_8}{8!}\phi_{q_1}\cdots \phi_{q_8}\hat{\delta}(q_1+\cdots +q_8)\nonumber\\
&&+\int _ {q_1 \cdots q_{10}}\frac{\lambda_{10}}{10!}\phi_{q_1}\cdots \phi_{q_{10}}\hat{\delta}(q_1+\cdots+q_{10}).\end{eqnarray}
Using this action, we calculate the first term on the RHS of the Wegner-Houghton equation \eref{eq00}
{\allowdisplaybreaks\begin{eqnarray}
M_{k', k}&\equiv&\langle{k'}|M|k\rangle
\equiv\frac{\delta^2S}{\delta \phi_{-k'}\delta \phi_{k}}\nonumber\\
&=&(k_0 ^2 +\alpha_0 k_i ^2 +\beta_0k_i ^2k_j ^2 +m^4)\hat{\delta}(k-k')\nonumber\\
&&+ \int _ {q_3 q_4} \frac{1}{2!}\Bigl\{ \alpha_2 \Bigl(q_{3, i} ^2+\frac{{k_i}^2}{2}+\frac{{k'_i} ^2}{2}\Bigl)+\lambda_4 \Bigl\} \phi_{q_3}\phi_{q_4}\hat{\delta}(k-k'+q_3+q_4)\nonumber\\
&&+ \int _ {q_3 \cdots q_6}\frac{1}{4!} \Bigl\{ \alpha_4 \Bigl(2{q_{3, i}}^2+\frac{{k_i}^2}{2}+\frac{{k'_i}^2}{2}\Bigl)+\lambda_6 \Bigl\} \phi_{q_3}\cdots \phi_{q_6}\hat{\delta}(k-k'+q_3+\cdots+q_6) \nonumber\\
&&+\int _ {q_3 \cdots q_8}\frac{\lambda_8}{6!} \phi_{q_3}\cdots \phi_{q_8}\hat{\delta}(k-k'+q_3+\cdots+q_8)\nonumber\\
&&+\int _ {q_3 \cdots q_{10}}\frac{\lambda_{10}}{8!} \phi_{q_3}\cdots \phi_{q_{10}}\hat{\delta}(k-k'+q_3+\cdots+q_{10}).
\end{eqnarray}}$M$ is split into the kinetic terms $A_0$, which involve the mass term, and the interaction terms $B$, 
\begin{eqnarray}
M=A_0+B.
\end{eqnarray}
Then, we obtain 
\begin{eqnarray}
\mathrm{tr}\ln(M)&=&\mathrm{tr}\ln(A_0+B)\nonumber\\
&=&\mathrm{tr}\ln(A_0)+\mathrm{tr}\sum_{n=1}^{\infty}\frac{(-1)^{n+1}}{n}(A_0^{-1}B)^n\nonumber\\
&=&\mathrm{tr}\ln(A_0)+\mathrm{tr}(A_0^{-1}B)-\frac{1}{2}\mathrm{tr}(A_0^{-1}B)^{2}+\cdots\nonumber\\
&\equiv&\raisebox{-2.5mm}{\includegraphics [width=6mm]{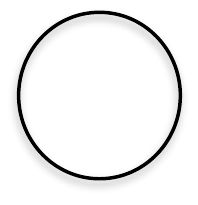}}+\raisebox{-2.5mm}{\includegraphics [width=6.5mm]{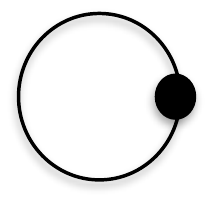}}-\frac{1}{2}\raisebox{-2.5mm}{\includegraphics [width=6.9mm]{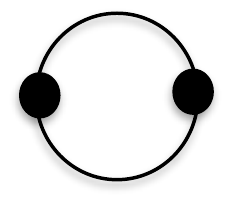}}+\cdots.\label{eq6000}
\end{eqnarray}
For example, the second and third terms in \eref{eq6000} are
{\allowdisplaybreaks \begin{eqnarray}
\mathrm{tr}(A_0^{-1}B)&=& \int_{k } \Bigl\{ \frac{1}{k_0 ^2 +\alpha_0 {k_i}^2 +\beta_0k_i ^2k_j ^2 +m^4}\Bigl\}\nonumber\\
&&\times\Bigl[\int _ {q_3 q_4} \frac{1}{2!}\{ \alpha_2 (q_{3, i} ^2+{k_i}^2)+\lambda_4 \} \phi_{q_3}\phi_{q_4}\hat{\delta}(q_3+q_4)\nonumber\\
&&+\int _ {q_3 \cdots q_6} \frac{1}{4!}\{ \alpha_4 (2q_{3, i} ^2+{k_i}^2)+\lambda_6\} \phi_{q_3}\cdots \phi_{q_6}\hat{\delta}(q_3+\cdots+q_6)\nonumber\\
&&+\int _ {q_3 \cdots q_8} \frac{1}{6!}\lambda_8  \phi_{q_3}\cdots \phi_{q_8}\hat{\delta}(q_3+\cdots+q_8)\nonumber\\
&&+\int _ {q_3 \cdots q_{10}} \frac{1}{8!}\lambda_{10}  \phi_{q_3}\cdots \phi_{q_{10}}\hat{\delta}(q_3+\cdots+q_{10})\Bigl],\end{eqnarray}}
{\allowdisplaybreaks \begin{eqnarray}
\lefteqn{ \mathrm{tr}(A_0^{-1}B)^2}\nonumber\\
&=& \int_{k  l^{(1)} } \Bigl\{ \frac{1}{k_0 ^2 +\alpha_0 {k_i}^2 +\beta_0k_i ^2k_j ^2 +m^4}\Bigl\}\nonumber\\
&&\times\Bigl[\int _ {q_3 q_4} \frac{1}{2!}\Bigl\{ \alpha_2 \Bigl(q_{3, i} ^2+\frac{{k_i}^2}{2}+\frac{(-l^{(1)}_i )^2}{2}\Bigl)+\lambda_4 \Bigl\} \phi_{q_3}\phi_{q_4}\hat{\delta}(k-l^{(1)}+q_3+q_4)\nonumber\\
&&+\int _ {q_3 \cdots q_6} \frac{1}{4!}\Bigl\{ \alpha_4 \Bigl(2q_{3, i} ^2+\frac{{k_i}^2}{2}+\frac{(-l^{(1)}_i )^2}{2}\Bigl)+\lambda_6\Bigl\} \phi_{q_3}\cdots \phi_{q_6}\hat{\delta}(k-l^{(1)}+q_3+\cdots+q_6)\nonumber\\
&&+\int _ {q_3 \cdots q_8} \frac{1}{6!}\lambda_8  \phi_{q_3}\cdots \phi_{q_8}\hat{\delta}(k-l^{(1)}+q_3+\cdots+q_8)\nonumber\\
&&+\int _ {q_3 \cdots q_{10}} \frac{1}{8!}\lambda_{10}  \phi_{q_3}\cdots \phi_{q_{10}}\hat{\delta}(k-l^{(1)}+q_3+\cdots+q_{10})\Bigl]\nonumber\\
&& \times\Bigl\{ \frac{1}{{l^{(1)}_0} ^2 +\alpha_0 {l^{(1)}_i} ^2 +\beta_0 {l^{(1)}_i} ^2 {l^{(1)}_j} ^2 +m^4}
\Bigl\} \nonumber\\
&&\times\Bigl[\int _ {q'_3 q'_4} \frac{1}{2!}\Bigl\{ \alpha_2 \Bigl({q'}_{3, i} ^2+\frac{{l^{(1)}_i}^2}{2}+\frac{(-k_i )^2}{2}\Bigl)+\lambda_4 \Bigl\} \phi_{q'_3}\phi_{q'_4}\hat{\delta}(l^{(1)}-k+q'_3+q'_4)\nonumber\\
&&+\int _ {q'_3 \cdots q'_6} \frac{1}{4!}\Bigl\{ \alpha_4 \Bigl(2{q'}_{3, i} ^2+\frac{{l^{(1)}_i}^2}{2}+\frac{(-k_i )^2}{2}\Bigl)+\lambda_6\Bigl\} \phi_{q'_3}\cdots \phi_{q'_6}\hat{\delta}(l^{(1)}-k+q'_3+\cdots+q'_6)\nonumber\\
&&+\int _ {q'_3 \cdots q'_8} \frac{1}{6!}\lambda_8  \phi_{q'_3}\cdots \phi_{q_8}\hat{\delta}(l^{(1)}-k+q'_3+\cdots+q'_8)\nonumber\\
&&+\int _ {q'_3 \cdots q'_{10}} \frac{1}{8!}\lambda_{10}  \phi_{q'_3}\cdots \phi_{q'_{10}}\hat{\delta}(l^{(1)}-k+q'_3+\cdots+q'_{10})\Bigl]\end{eqnarray}}and so on. In order to calculate Eq.~\eref{eq6000} systematically through the order that we need in the model, we define $G$ as an integrand of elements of the kinetic term $A_0^{-1}$ and $V$ as an integrand of elements of interaction terms $B$. We also define $G_n$ and $V_n$ as terms that have $n$ power of external momentum in $G$ and $V$.  We call $n$ the ``derivative number'' in this paper.  Now, we need only terms $n\leq2$; we write
\begin{eqnarray}
G&=&G_0+G_1+G_2,\\
V&=&V_0+V_1+V_2,
\end{eqnarray}
where
{\allowdisplaybreaks \begin{eqnarray}
G&=&\frac{1}{(k_0+r_0) ^2 +\alpha_0 (k_i+r_i)^2 +\beta_0(k_i +r_i)^2(k_j+r_j) ^2 +m^4},\\
G_0&=&\frac{1}{k_0 ^2 +\alpha_0 k_i ^2 +\beta_0k_i ^2k_j ^2 +m^4},\\
G_1&=&-{G_0}^2 (\alpha_0  +2\beta _0 k_j^2 )2 k_ir_i,\\
G_2&=&-{G_0}^2 \Bigl(\alpha_0 +2\beta_0k_j^2+\frac{4\beta_0}{D}k_j^2\Bigl) r_i^2+G_0^3(\alpha_0+2\beta_0k_j)^2 \frac{4k_k^2 r_i^2}{D},
\end{eqnarray}}
and
{\allowdisplaybreaks \begin{eqnarray}
V_0&=&\frac{1}{2!}(\alpha_2 k_i^2 +\lambda_4)\phi^2+\frac{1}{4!}(\alpha_4 k_i^2 +\lambda_6)\phi^4+\frac{1}{6!}\lambda_8\phi^6+\frac{1}{10!}\lambda_{10}\phi^8,\\
V_1&=&\Bigl(\frac{1}{2!}\alpha_2 \phi^2+\frac{1}{4!}\alpha_4\phi^4\Bigl)(k_i r_i+k_j r'_j),\\
V_2&=&\frac{1}{2!}\alpha_2\Bigl(q_{3, i}^2 +\frac{r_i^2+{r'}_j^2}{2} \Bigl) \phi^2+\frac{1}{4!}\alpha_4\Bigl(2q_{3, i}^2 +\frac{r_i^2+{r'}_j^2}{2}\Bigl)\phi^4,
\end{eqnarray}}where $r$ and $r'$ are particular sums of external momenta $q_i$'s, which are obtained after integrating out all the internal momenta except $k$. 
 Omitting the part about the delta function, we write each term of Eq.~\eref{eq6000} symbolically as
{\allowdisplaybreaks \begin{eqnarray}
\mathrm{tr}(A_0^{-1}B)&\leftrightarrow& \int_{k}G_0 (V_0+V_1+V_2),\\
\mathrm{tr}(A_0^{-1}B)^2&\leftrightarrow& \int_{k}G_0(V_0+V_1+V_2)(G_0+G_1+G_2)(V_0+V_1+V_2),\hspace{2cm}\\
\mathrm{tr}(A_0^{-1}B)^3&\leftrightarrow&\int_{k}G_0(V_0+V_1+V_2)\{(G_0+G_1+G_2)(V_0+V_1+V_2)\}^2,\\
\mathrm{tr}(A_0^{-1}B)^4&\leftrightarrow&\int_{k}G_0(V_0+V_1+V_2)\{(G_0+G_1+G_2)(V_0+V_1+V_2)\}^3,\\
\mathrm{tr}(A_0^{-1}B)^5&\leftrightarrow&\int_{k}G_0(V_0+V_1+V_2)\{(G_0+G_1+G_2)(V_0+V_1+V_2)\}^4,
\end{eqnarray}}where we choose integration variables as $G_0$ in front of the integrand.  We list the integrand as follows in Table~\ref{table1}

\begin{table}[t]
\caption{Classification of the integrand. $k$ is the order of the polynomial expansion of $\mathrm{tr}\ln(1+A_0^{-1}B)$, and n is the derivative number.}\label{table1}
\[
\begin{array}{ccccccc}\hline\hline
k=1, n=2&k=2, n=2&k=3, n=2&k=4, n=2&k=5, n=2\\\\
G_0 V_2\cdots(1)& G_0 V_2 G_0 V_0\cdots(2)  &G_0 V_2 G_0 V_0  G_0 V_0\cdots(8) &{\rm nothing}&{\rm nothing}\\
&G_0 V_0 G_0 V_2\cdots(3) & G_0 V_0 G_0V_2  G_0V_0\cdots(9)\\
& G_0 V_1G_0 V_1\cdots(4) & G_0 V_0 G_0V_0 G_0V_2\cdots(10)\\
 &G_0 V_1G_1 V_0\cdots(5) & G_0 V_1G_0 V_1 G_0V_0\cdots(11)\\
 &G_0 V_0 G_1 V_1\cdots(6) & G_0 V_1  G_0V_0 G_0V_1\cdots(12)\\
&G_0 V_0 G_2V_0\cdots(7)& G_0 V_0 G_0V_1 G_0V_1\cdots(13)\\
&&G_0 V_1 G_0V_0  G_1V_0\cdots(14)\\
&&G_0 V_1 G_1V_0 G_0V_0\cdots(15)\\
&&G_0 V_0 G_0V_1  G_1V_0\cdots(16)\\
&&G_0 V_0  G_1V_1 G_0V_0\cdots(17)\\
&&G_0 V_0  G_0V_0 G_1V_1\cdots(18)\\
&& G_0 V_0 G_1V_0G_0 V_1\cdots(19)\\
&& G_0 V_0 G_1V_ 0  G_1V_0\cdots(20)\\
 &&G_0 V_0  G_0V_0 G_2V_0\cdots(21)\\
 &&G_0 V_0 G_2V_0 G_0V_0 \cdots(22)\\
\end{array}\]\[
\begin{array}{ccccccc}
k=1, n=0&k=2, n=0&k=3,n=0&k=4, n=0&k=5, n=0\\\\
G_0 V_0\cdots(23)&(G_0 V_0)^2 \cdots(24)  &(G_0 V_0)^3\cdots(25) &(G_0 V_0)^4\cdots(26)&(G_0 V_0)^5\cdots(27)\\
\hline\end{array}
\]

\end{table}

We give the results of the integration for (1) -- (27) in Appendix \ref{appd}. In this paper, we integrate out in a timelike direction from $-\infty$ to $\infty$, and spacelike directions within shell-mode momentum $1-\delta t \leq \mid k_i \mid  \leq 1$.  That is, we consider only the spatial flow of the theory. 
 \begin{eqnarray}
\int_{k}G_0 &=&\int_{-\infty}^{\infty}\frac{dk_0}{2\pi}\int_{1-\delta t \leq \mid k_i \mid  \leq 1}\frac{d^{d-1}k_i}{(2\pi)^{d-1}}
\frac{1}{k_0^2+\alpha_0 k_i^2+\beta_0 k_i^2 k_j^2+m^4}\nonumber\\
&=&\frac{1}{2}\int_{1-\delta t \leq \mid k_i \mid  \leq 1}\frac{d^{d-1}k_i}{(2\pi)^{d-1}}\frac{1}{(\alpha_0 k_i^2+\beta_0 k_i^2 k_j^2+m^4)^{1/2}}\nonumber\\
&=&\frac{1}{2}A_{d-1}\frac{\delta t}{(\alpha_0 +\beta_0+m^4) ^{1/2}},
\end{eqnarray}
where $A_{d-1}$ is a superficial area of a unit sphere in $d-1$ space-time dimensions divided by $(2\pi)^{d-1}$.  In general, the result is the same even if the numerator of the integrand has any $2n$ power of spatial momentum. We obtain the formula given as
{\allowdisplaybreaks\begin{eqnarray}
\int_{k}G_0 k_i^{2n}
&=&\frac{1}{2}A_{d-1}\frac{\delta t}{(\alpha_0 +\beta_0+m^4) ^{1/2}},\\
\int_{k}G_0^2 k_i^{2n}
&=&\frac{1}{4}A_{d-1}\frac{\delta t}{(\alpha_0 +\beta_0+m^4) ^{3/2}},\\
\int_{k}G_0^3 k_i^{2n}
&=&\frac{3}{16}A_{d-1}\frac{\delta t}{(\alpha_0 +\beta_0+m^4) ^{5/2}},\\
\int_{k}G_0^4 k_i^{2n}
&=&\frac{5}{32}A_{d-1}\frac{\delta t}{(\alpha_0 +\beta_0+m^4) ^{7/2}},\\
\int_{k}G_0^5 k_i^{2n}
&=&\frac{35}{256}A_{d-1}\frac{\delta t}{(\alpha_0 +\beta_0+m^4) ^{9/2}}.
\end{eqnarray}}

The second term on the RHS of the Wegner-Houghton equation \eref{eq00} is 0 in the case that the external momentum is smaller than the shell-mode momentum.  The second and third lines of the Wegner-Houghton equation \eref{eq00} are
\begin{eqnarray}
\lefteqn{-(D+z)S}\nonumber\\
&=&\int_{x}-\frac{5}{2}(-\phi\partial_0\partial_0\phi-\alpha_0\phi\partial_i \partial_i \phi+\beta_0\phi\partial_i \partial_i\partial_j \partial_j\phi+m^4\phi^2)\nonumber\\
&&+5\int_x\frac{1}{12}\alpha_2\phi^3\partial_i \partial_i \phi+5\int_x\frac{1}{240}\alpha_4\phi^5\partial_i \partial_i \phi\nonumber\\
&&-\frac{5}{4!}\lambda_4\int_x\phi^4-\frac{5}{6!}\lambda_6\int_x\phi^6-\frac{5}{8!}\lambda_8\int_x\phi^8-\frac{5}{10!}\lambda_{10}\int_x\phi^{10},
\end{eqnarray}
and
{\allowdisplaybreaks \begin{eqnarray}
\lefteqn{\int_{p}\Omega_{p}^{i}\Bigl(d_{\phi}-\gamma+\hat{p}^{\mu} \frac{\partial'}{\partial \hat{p}^{\mu}}\Bigl) \frac{\delta}{\delta \Omega_{p}^{i}}S}\nonumber\\
&=&\int_{x}\Bigl(\frac{1}{2}-\gamma\Bigl)\Bigl(-\phi\partial_0\partial_0\phi-\alpha_0\phi\partial_i \partial_i \phi+\beta_0\phi\partial_i \partial_i\partial_j \partial_j\phi \nonumber\\
&&+m^4\phi^2+\frac{1}{3!}\lambda_4\phi^4+\frac{1}{5!}\lambda_6\phi^6+\frac{1}{7!}\lambda_8\phi^8+\frac{1}{9!}\lambda_{10}\phi^{10}\nonumber\\
&&-\frac{1}{3}\alpha_2\phi^3\partial_i \partial_i \phi-\frac{1}{40}\alpha_4\phi^5\partial_i \partial_i \phi\Bigl)\nonumber\\
&&+\int_{x}(-2\phi\partial_0\partial_0\phi-\alpha_0\phi\partial_i \partial_i \phi+2\phi\beta_0\partial_i\partial_i\partial_j\partial_j\phi)\nonumber\\
&&+\int_x\Bigl(-\frac{1}{6}\alpha_2\phi^3\partial_i \partial_i\phi\Bigl)+\int_x \Bigl(-\frac{1}{120}\alpha_4\phi^5\partial_i \partial_i\phi\Bigl).
\end{eqnarray}}
It is found from the derivative term with respect to time that there is an anomalous dimension $\gamma=0$. Finally, we obtain Eqs.~\eref{eq1}--\eref{eq2}.

\section{The Results of (1)--(27)}\label{appd}
Here, we present integral relations for Eqs.~(1)--(27).
{\allowdisplaybreaks

\begin{eqnarray*}
(1)-\int_{x}\phi \partial_i\partial_i\phi  \hspace{2mm}& \leftrightarrow&\int_{k}G_0\frac{\alpha_2}{2!}. \hspace{9cm}\\
     -\int_{x}\phi^3 \partial_i\partial_i\phi  \hspace{2mm}& \leftrightarrow&\int_{k}G_0\frac{\alpha_4}{12}.\\
(2)-\int_{x}\phi^3 \partial_i\partial_i\phi \hspace{2mm} &\leftrightarrow&\int_{k}G_0^2\frac{5}{12}\alpha_2(\alpha_2 k_{i}^2+\lambda_4 ).\\
     -\int_{x}\phi^5 \partial_i\partial_i\phi  \hspace{2mm}& \leftrightarrow&\int_{k}G_0^2\Bigl(\frac{23}{240}\alpha_2\alpha_4 k_{i}^2+\frac{14}{240}\alpha_4\lambda_4+\frac{9}{240}\alpha_2\lambda_6\Bigl).\\
 (3)-\int_{x}\phi^3 \partial_i\partial_i\phi  \hspace{2mm} &\leftrightarrow&\int_{k}G_0^2\frac{5}{12}\alpha_2(\alpha_2 k_{i}^2+\lambda_4 ).\\
     -\int_{x}\phi^5 \partial_i\partial_i\phi  \hspace{2mm} &\leftrightarrow&\int_{k}G_0^2\Bigl(\frac{23}{240}\alpha_2\alpha_4 k_{i}^2+\frac{14}{240}\alpha_4\lambda_4+\frac{9}{240}\alpha_2\lambda_6\Bigl).\\
     (4)-\int_{x}\phi^3 \partial_i\partial_i\phi  \hspace{2mm} &\leftrightarrow&\int_{k}G_0^2\frac{1}{3D}k_{i}^2 \alpha_2^2.\\
          -\int_{x}\phi^5 \partial_i\partial_i\phi  \hspace{2mm} &\leftrightarrow&\int_{k}G_0^2\frac{1}{15D}k_{i}^2 \alpha_2 \alpha_4.\\
     (5)-\int_{x}\phi^3 \partial_i\partial_i\phi  \hspace{2mm} &\leftrightarrow&  \int_{k}G_0^3\frac{-2}{3D}k_{i}^2 \alpha_2 (\alpha_2 k_{i}^2+\lambda_4 )(\alpha_0+2\beta_0 k_{i}^2).\\
          -\int_{x}\phi^5 \partial_i\partial_i\phi  \hspace{2mm} &\leftrightarrow& \int_{k}G_0^3\frac{-1}{15D}k_{i}^2 (\alpha_0+2\beta_0 k_{i}^2)(2\alpha_2\alpha_4 k_{i}^2+\alpha_4\lambda_4+\alpha_2\lambda_6  ).\\
     (6)-\int_{x}\phi^3 \partial_i\partial_i\phi  \hspace{2mm} &\leftrightarrow&  \int_{k}G_0^3\frac{-2}{3D}k_{i}^2 \alpha_2 (\alpha_2 k_{i}^2+\lambda_4 )(\alpha_0+2\beta_0 k_{i}^2).\\
          -\int_{x}\phi^5 \partial_i\partial_i\phi  \hspace{2mm} &\leftrightarrow& \int_{k}G_0^3\frac{-1}{15D}k_{i}^2 (\alpha_0+2\beta_0 k_{i}^2)(2\alpha_2\alpha_4 k_{i}^2+\alpha_4\lambda_4+\alpha_2\lambda_6  ).\\
     (7)-\int_{x}\phi^3 \partial_i\partial_i\phi  \hspace{2mm} &\leftrightarrow&  \int_{k}G_0^3\frac{-1}{3}(\alpha_2 k_{i}^2+\lambda_4 )^2\Bigl(\alpha_0+2\beta_0 k_{i}^2+4\frac{\beta_0 k_{i}^2}{D}\Bigl) \nonumber\\
     &&+\int_{k}G_0^4\frac{4}{3D}k_{i}^2 (\alpha_2 k_{i}^2+\lambda_4 )^2(\alpha_0+2\beta_0 k_{i}^2)^2.\\
         -\int_{x}\phi^5 \partial_i\partial_i\phi  \hspace{2mm} &\leftrightarrow&  \int_{k}G_0^3\frac{-1}{15} (\alpha_2 k_{i}^2+\lambda_4 )(\alpha_4 k_{i}^2+\lambda_6 )\Bigl(\alpha_0+2\beta_0 k_{i}^2+4\frac{\beta_0 k_{i}^2}{D}\Bigl) \nonumber\\
         &&+\int_{k}G_0^4\frac{4}{15D}k_{i}^2(\alpha_0+2\beta_0 k_{i}^2)^2  (\alpha_2 k_{i}^2+\lambda_4 )(\alpha_4 k_{i}^2+\lambda_6 ).\\
 (8) -\int_{x}\phi^5 \partial_i\partial_i\phi  \hspace{2mm} &\leftrightarrow&\int_{k}G_0^3\frac{9}{40}\alpha_2(\alpha_2 k_{i}^2+\lambda_4)^2.\\
 (9) -\int_{x}\phi^5 \partial_i\partial_i\phi  \hspace{2mm} &\leftrightarrow&\int_{k}G_0^3\frac{13}{40}\alpha_2(\alpha_2 k_{i}^2+\lambda_4)^2.\\
 (10) -\int_{x}\phi^5 \partial_i\partial_i\phi  \hspace{2mm} &\leftrightarrow&\int_{k}G_0^3\frac{9}{40}\alpha_2(\alpha_2 k_{i}^2+\lambda_4)^2.\\
  (11) -\int_{x}\phi^5 \partial_i\partial_i\phi  \hspace{2mm} &\leftrightarrow&\int_{k}G_0^3\frac{3}{10D}k_{i}^2\alpha_2^2(\alpha_2 k_{i}^2+\lambda_4).\\
  (12) -\int_{x}\phi^5 \partial_i\partial_i\phi  \hspace{2mm} &\leftrightarrow&\int_{k}G_0^3\frac{1}{10D}k_{i}^2\alpha_2^2(\alpha_2 k_{i}^2+\lambda_4).\\
  (13) -\int_{x}\phi^5 \partial_i\partial_i\phi  \hspace{2mm} &\leftrightarrow&\int_{k}G_0^3\frac{3}{10D}k_{i}^2\alpha_2^2(\alpha_2 k_{i}^2+\lambda_4). \\
    (14) -\int_{x}\phi^5 \partial_i\partial_i\phi  \hspace{2mm} &\leftrightarrow&\int_{k}G_0^4\frac{-1}{5D}k_{i}^2\alpha_2(\alpha_2 k_{i}^2+\lambda_4)^2 (\alpha_0+2\beta_0 k_{i}^2).\\
  (15) -\int_{x}\phi^5 \partial_i\partial_i\phi  \hspace{2mm} &\leftrightarrow&\int_{k}G_0^4\frac{-2}{5D}k_{i}^2\alpha_2(\alpha_2 k_{i}^2+\lambda_4)^2 (\alpha_0+2\beta_0 k_{i}^2).\\
  (16) -\int_{x}\phi^5 \partial_i\partial_i\phi  \hspace{2mm} &\leftrightarrow&\int_{k}G_0^4\frac{-3}{5D}k_{i}^2\alpha_2(\alpha_2 k_{i}^2+\lambda_4)^2 (\alpha_0+2\beta_0 k_{i}^2).\\
   (17) -\int_{x}\phi^5 \partial_i\partial_i\phi  \hspace{2mm} &\leftrightarrow&\int_{k}G_0^4\frac{-3}{5D}k_{i}^2\alpha_2(\alpha_2 k_{i}^2+\lambda_4)^2 (\alpha_0+2\beta_0 k_{i}^2).\\
    (18) -\int_{x}\phi^5 \partial_i\partial_i\phi  \hspace{2mm} &\leftrightarrow&\int_{k}G_0^4\frac{-2}{5D}k_{i}^2\alpha_2(\alpha_2 k_{i}^2+\lambda_4)^2 (\alpha_0+2\beta_0 k_{i}^2).\\
     (19) -\int_{x}\phi^5 \partial_i\partial_i\phi  \hspace{2mm}& \leftrightarrow&\int_{k}G_0^4\frac{-1}{5D}k_{i}^2\alpha_2(\alpha_2 k_{i}^2+\lambda_4)^2 (\alpha_0+2\beta_0 k_{i}^2).\\
      (20) -\int_{x}\phi^5 \partial_i\partial_i\phi  \hspace{2mm} &\leftrightarrow&\int_{k}G_0^5\frac{2}{5D}k_{i}^2(\alpha_2 k_{i}^2+\lambda_4)^3 (\alpha_0+2\beta_0 k_{i}^2)^2.\\
       (21) -\int_{x}\phi^5 \partial_i\partial_i\phi  \hspace{2mm} &\leftrightarrow &\int_{k}G_0^4\frac{-1}{5}(\alpha_2 k_{i}^2+\lambda_4)^3\Bigl(\alpha_0+2\beta_0 k_{i}^2+4\beta_0\frac{k_{i}^2}{D}\Bigl)\nonumber\\
       &&+\int_{k}G_0^5\frac{4}{5D}k_{i}^2(\alpha_2 k_{i}^2+\lambda_4)^3(\alpha_0+2\beta_0 k_{i}^2)^2.\\
         (22) -\int_{x}\phi^5 \partial_i\partial_i\phi  \hspace{2mm} &\leftrightarrow& \int_{k}G_0^4\frac{-1}{5}(\alpha_2 k_{i}^2+\lambda_4)^3\Bigl(\alpha_0+2\beta_0 k_{i}^2+4\beta_0\frac{k_{i}^2}{D}\Bigl)\nonumber\\
       &&+\int_{k}G_0^5\frac{4}{5D}k_{i}^2(\alpha_2 k_{i}^2+\lambda_4)^3(\alpha_0+2\beta_0 k_{i}^2)^2.\\
(23)\hspace{7mm}\int_{x}\phi^2    \hspace{7mm} &\leftrightarrow&\int_{k}G_0\frac{1}{2!}(\alpha_2 k_{i}^2+\lambda_4 ).\\
     \int_{x}\phi^4  \hspace{7mm} &\leftrightarrow&\int_{k} G_0\frac{1}{4!}(\alpha_4 k_{i}^2+\lambda_6 ).\\
     \int_{x}\phi^6\hspace{7mm} &\leftrightarrow&\int_{k}G_0\frac{1}{6!}\lambda_8. \\
     \int_{x}\phi^8\hspace{7mm} &\leftrightarrow&\int_{k}G_0\frac{1}{8!}\lambda_{10}. \\
(24)\hspace{7mm}\int_{x}\phi^4   \hspace{7mm} &\leftrightarrow&\int_{k}G_0^2\frac{1}{4}(\alpha_2 k_{i}^2+\lambda_4 )^2.\\
       \int_{x}\phi^6 \hspace{7mm} &\leftrightarrow&\int_{k}G_0^2\frac{1}{4!}(\alpha_2 k_{i}^2+\lambda_4 ) (\alpha_4 k_{i}^2+\lambda_6).\\
       \int_{x}\phi^8 \hspace{7mm}& \leftrightarrow&\int_{k}G_0^2 \Bigl\{\frac{1}{6!}(\alpha_2 k_{i}^2+\lambda_4 )\lambda_8+\frac{1}{4!^2}(\alpha_4 k_{i}^2+\lambda_6)^2\Bigl\}.\\
       \int_{x}\phi^{10} \hspace{7mm}& \leftrightarrow&\int_{k}G_0^2\Bigl\{\frac{2}{4!\times6!}(\alpha_4 k_{i}^2+\lambda_6 )\lambda_8+\frac{1}{8!}(\alpha_2 k_{i}^2+\lambda_4)\lambda_{10}\Bigl\}.\\
(25)\hspace{7mm} \int_{x}\phi^6 \hspace{7mm}& \leftrightarrow&\int_{k}G_0^3\frac{1}{8}(\alpha_2 k_{i}^2+\lambda_4)^3. \\
       \int_{x}\phi^8\hspace{7mm} &\leftrightarrow&\int_{k}G_0^3\frac{3}{4\times4!}(\alpha_2 k_{i}^2+\lambda_4)^2 (\alpha_4 k_{i}^2+\lambda_6).  \\
       \int_{x}\phi^{10} \hspace{7mm} &\leftrightarrow&\int_{k}G_0^3\Bigl\{ \frac{3}{2\times 4!^2}(\alpha_2 k_{i}^2+\lambda_4) (\alpha_4 k_{i}^2+\lambda_6)^2 +\frac{3}{4 \times6!}(\alpha_2 k_{i}^2+\lambda_4)^2 \lambda_8  \Bigl\}.    \\
(26)\hspace{7mm} \int_{x}\phi^8 \hspace{7mm} &\leftrightarrow&\int_{k}G_0^4\frac{1}{2^4}(\alpha_2 k_{i}^2+\lambda_4)^4.\\
       \int_{x}\phi^{10} \hspace{7mm} &\leftrightarrow&\int_{k}G_0^4\frac{1}{2^4\times3}(\alpha_2 k_{i}^2+\lambda_4)^3    (\alpha_4 k_{i}^2+\lambda_6).  \\
(27) \hspace{6mm} \int_{x}\phi^{10} \hspace{7mm} &\leftrightarrow&\int_{k}G_0^5\frac{1}{2^5}(\alpha_2 k_{i}^2+\lambda_4)^5.    
\end{eqnarray*}}

\providecommand{\href}[2]{#2}\begingroup\raggedright\endgroup

\end{document}